\documentstyle[12pt,epsfig]{article}
\renewcommand{\theequation}{\thesection.\arabic{equation}}
\newcommand{\beq}{\begin{equation}}
\newcommand{\eeq}{\end{equation}}
\newcommand{\bea}{\begin{eqnarray}}
\newcommand{\eea}{\end{eqnarray}}
\newcommand{\gsim}{\raisebox{-0.07cm}{$\:\stackrel{>}{{\scriptstyle
 \sim}}\: $} }
\newcommand{\lsim}{\raisebox{-0.07cm}{$\:\stackrel{<}{{\scriptstyle
 \sim}}\: $} }
\newcommand{\ra}{\!\rightarrow\!}
\newcommand\cV{\mbox{\boldmath $c$}}
\newcommand\qV{\mbox{\boldmath $q$}}
\newcommand\CV{\mbox{\boldmath $C$}}
\newcommand\PV{\mbox{\boldmath $P$}}
\newcommand\ZV{\mbox{\boldmath $Z$}}
\newcommand\eV{\mbox{\boldmath $1$}}
\newcommand\MSb{$\overline{\mbox{MS}}$}
\begin{document}
\setlength{\parskip}{0.5cm}
\setlength{\baselineskip}{0.6cm}
\begin{titlepage}

\noindent
{\tt hep-ph/0006154} \hfill INLO-PUB 04/00 \\ 
\hspace*{\fill} May 2000 \\
\vspace{1.0cm}
\begin{center}
\Large
{\bf NNLO Evolution of Deep-Inelastic} \\
\vspace{0.15cm}
{\bf Structure Functions: the Singlet Case} \\
\vspace{2.2cm}
\large
W.L. van Neerven and A. Vogt \\
\vspace{0.8cm}
\normalsize
{\it Instituut-Lorentz, University of Leiden \\
\vspace{0.1cm}
P.O. Box 9506, 2300 RA Leiden, The Netherlands} \\
\vspace{3.0cm}
{\bf Abstract}
\vspace{-0.3cm}
\end{center}
We study the next-to-next-to-leading order (NNLO) evolution of flavour
singlet parton densities and structure functions in massless 
perturbative QCD.
Present information on the corresponding three-loop splitting functions
is used to derive parame\-trizations of these quantities, including
Bjorken-$x$ dependent estimates of their residual uncertainties.
Compact expressions are also provided for the exactly known, but in 
part rather lengthy two-loop singlet coefficient functions.
The size of the NNLO corrections and their effect on the stability
under variations of the renormalization and mass-factorizations scales
are investigated. Except for rather low values of the scales, the 
residual 
uncertainty of the three-loop splitting functions does not lead to 
relevant effects for $x>10^{-3}$.  Inclusion of the NNLO contributions 
considerably reduces the theoretical uncertainty of determinations of 
the quark and gluon densities from deep-inelastic structure functions.

\vspace{0.3cm}
\begin{center}
\end{center}
\vspace{-0.3cm}

\end{titlepage}
%
%
\section{Introduction}
%
%
Structure functions in deep-inelastic scattering (DIS) form the 
backbone of our \mbox{knowledge} of the proton's parton densities --- 
which are indispensable for analyses of hard scattering processes at 
proton--(anti-)proton colliders like {\sc Tevatron} and the future LHC 
--- and are among the quantities best suited for measuring the strong
coupling constant $\alpha_s$. The full realization of this potential,
however, requires transcending the standard next-to-leading order (NLO)
approximation of perturbative QCD summarized in ref.~\cite{FP82}.
In fact, the next-to-next-to-leading order (NNLO) subprocess 
cross-sections (`coefficient functions') for DIS have been calculated 
some time ago \cite{ZvN1,ZvN2} (see also refs.~\cite{c2DY} for the 
related Drell-Yan process), but the corresponding three-loop splitting 
functions governing the NNLO scale dependence (`evolution') of the
parton densities have not been completed so far (see ref.~\cite{MV2} 
for an up-to-date progress report).

In a previous paper \cite{NV1} we have studied the flavour non-singlet
sector of the DIS structure functions. It turned out that the available 
incomplete information on the non-singlet splitting functions
\cite{spfm1,spfm2,Gra1,BVns} facilitates the derivation of approximate 
expressions which are sufficiently accurate over a rather wide region
of parton momenta and the Bjorken variable $x$. In the present 
paper, we complement those results by corresponding effective 
parametrizations for the flavour-singlet sector based on the partial 
information of refs.~\cite{spfm2,CH94,FL98,Gra2}, thus paving the way 
for promoting, even though only at $x\! >\! 10^{-3}$, global analyses 
\cite{MRS,CTEQ,GRV} of DIS and related processes \cite{c2DY} to NNLO 
accuracy.

This paper is built up as follows: 
In Sect.~2 we set up our notations by recalling the general framework
for the evolution of singlet parton densities and structure functions 
in massless perturbative QCD.  The expansions of the corresponding 
splitting functions and coefficient functions are written down up to 
order $\alpha_s^3$ for arbitrary values of the renor\-mali\-zation and 
mass-factorization scales.  
In Sect.~3 we present compact, but very accurate parametrizations of 
the exactly known \cite{c2Sa,ZvN1,ZvN2}, but partly rather lengthy
expressions for the two-loop singlet coefficient functions. 
In Sect.~4 the available constraints \cite{spfm2,CH94,FL98,Gra2} on the 
three-loop singlet splitting functions are utilized for constructing 
approximate expressions for the $x$-dependence of these functions, 
including quantitative estimates of the remaining uncertainties.  
In Sect.~5 we assemble all these results and quantify the impact of 
the NNLO contributions on the evolution of the singlet quark and gluon
densities and on the most important singlet structure function, $F_2$.
We address the range of applicability of the present approximate
results and the improvement of the theoretical accuracy of
determinations of the parton densities at NNLO.
Finally our results are summarized in Sect.~6.  Mellin-$N$ space 
expressions for our parametrizations of the two-loop coefficient 
functions of Sect.~3 can be found in the appendix.
%
%
\section{General Framework}
%
%
We start by outlining the general formalism for the NNLO evolution of
flavour-singlet parton densities and structure functions.  The singlet 
quark density of a hadron is given~by
\beq
  \Sigma(x,\mu_f^2,\mu_r^2) \: = \: \sum_{i=1}^{N_f} \left[ 
  q_i(x,\mu_f^2,\mu_r^2) + \bar{q}_i(x,\mu_f^2,\mu_r^2) \right] \:\: .
\eeq
Here $q_i(x,\mu_f^2,\mu_r^2)$ and $\bar{q}_i(x,\mu_f^2,\mu_r^2)$ 
represent the number distributions of quarks and anti\-quarks,
respectively, in the fractional hadron momentum $x$.  The corresponding
gluon distribution is denoted by $g(x,\mu_f^2,\mu_r^2)$.  The subscript 
$i$ indicates the flavour of the (anti-) quarks, and $N_f$ stands for 
the number of effectively massless flavours.  Finally $\mu_r$ and
$\mu_f$ represent the renormalization and mass-factorization scales, 
respectively.  The singlet quark density (2.1) and the gluon density 
are constrained by the energy-momentum sum rule
\beq
  \int_0^1 \! dx \: x \left[ \Sigma(x,\mu_f^2,\mu_r^2) + 
  g(x,\mu_f^2,\mu_r^2) ) \right] \: = \: 1 \:\: .
\eeq

The scale dependence of the singlet parton densities is given by the 
evolution equations
\beq
  \frac{d \qV}{d \ln \mu_f^2} \:\equiv\: \frac{d}{d \ln \mu_f^2}
  \left( \begin{array}{c} \!\Sigma\! \\ g  \end{array} \right) 
  \: = \: \left( \begin{array}{cc} {\cal P}_{qq} & {\cal P}_{qg} \\ 
  {\cal P}_{gq} & {\cal P}_{gg} \end{array} \right) \otimes 
  \left( \begin{array}{c} \!\Sigma\! \\ g  \end{array} \right) 
  \:\equiv\: \PV \otimes \qV \:\: ,
\eeq
where $\otimes$ stands for the Mellin convolution in the momentum
variable,
\beq
  [ a \otimes b ](x) \:\equiv\: \int_x^1 \! \frac{dy}{y} \: a(y)\,
  b\bigg(\frac{x}{y}\bigg) \:\: .
\eeq
As in some other equations below, the dependence on $x$, $\mu_f$ and
$\mu_r$ has been suppressed in Eq.~(2.3).  The splitting function 
${\cal P}_{qq}$ can be expressed as 
\beq
  {\cal P}_{qq} \: =\: {\cal P}_{\rm NS}^+ + {\cal P}_{\rm PS} 
\eeq
with
\beq
  {\cal P}_{\rm PS}^{\,} \: =\: N_f\, ({\cal P}_{qq}^S 
  + {\cal P}_{\bar{q}q}^S) \:\: .
\eeq
Here ${\cal P}_{\rm NS}^+$ is the non-singlet splitting function 
discussed up to NNLO in ref.~\cite{NV1}, and the ${\cal O}(\alpha_s^2)$
quantities ${\cal P}_{qq}^S$ and ${\cal P}_{\bar{q}q}^S$ are the 
flavour independent (`sea') contributions to the quark-\-quark and 
quark-antiquark splitting functions ${\cal P}_{q_{i}q_{k}}$ and 
${\cal P}_{\bar{q}_{i} q_{k}}$, respectively.  The non-singlet 
contribution dominates Eq.~(2.5) at large $x$, where the pure singlet 
term ${\cal P}_{\rm PS}^{\,}$ is very small.  At small $x$, on the 
other hand, the latter contribution takes over as 
$x{\cal P}_{\rm PS}^{\,}$, unlike $x{\cal P}_{\rm NS}^+ $, does not 
vanish for $x \!\rightarrow\! 0$.  The gluon-quark and quark-gluon 
entries in (2.3) are given by
\beq
  {\cal P}_{qg} \: =\: N_f\, {\cal P}_{q_{i}g} \:\: , \quad
  {\cal P}_{gq} \: =\: {\cal P}_{gq_{i}}
\eeq
in terms of the flavour-independent splitting functions 
${\cal P}_{q_{i}g} = {\cal P}_{\bar{q}_{i}g}$ and ${\cal P}_{gq_{i}} = 
{\cal P}_{g\bar{q}_{i}}$.  With the exception of the lowest-order
approximation to ${\cal P}_{qg}$, neither of the quantities 
$x{\cal P}_{qg}$, $x{\cal P}_{gq}$ and $x{\cal P}_{gg}$ vanishes for
$x \!\rightarrow\! 0$.  
 
\noindent
The NNLO expansion of the splitting-function matrix $\PV$ in Eq.~(2.3)
reads\footnote
{Notice that the convention adopted for $P^{(l)}$ here and in ref.\ 
 \cite{NV1} differs from ref.~\cite{FP82} by a factor of $2^{l+1}$ due 
 to the choice of $a_s = \alpha_s/(4\pi)$ instead of $\alpha_s/(2\pi)$ 
 as the expansion parameter in Eq.~(2.8), and by a factor 1/2 from 
 ref.~\cite{ZvN2} due to the choice of $\mu_f^2$ instead of $\mu_f$ for 
 the differentiation in Eq.~(2.3).}
\bea
  \PV \Big( x, \alpha_s(\mu_r^2), L_R \Big) &\! =\!\! & \quad
      a_s \, \PV^{(0)}(x) \nonumber \\
  & & \mbox{}\!\!
    + a_s^2 \, \Big( \PV^{(1)}(x) - \beta_0 L_R \PV^{(0)}(x) \Big) 
  \:  \\ & & \mbox{}\!\!
    + a_s^3 \, \Big( \PV^{(2)}(x) - 2\beta_0 L_R \PV^{(1)}(x) 
    - \Big\{ \beta_1 L_R - \beta_0^2 L_R^2 \Big\} \PV^{(0)}(x) \Big) 
    + \ldots \nonumber \:\: .
\eea
In Eq.~(2.8) and in what follows we use the abbreviations 
\beq
  a_s \:\equiv\: \frac{\alpha_s(\mu_r^2)}{4\pi} 
\eeq
for the running coupling, and 
\beq
  L_M \:\equiv\: \ln \frac{Q^2}{\mu_f^2} \:\: , \quad
  L_R \:\equiv\: \ln \frac{\mu_f^2}{\mu_r^2} 
\eeq
for the scale logarithms.  The one- and two-loop matrices 
$\PV^{(0)}(x)$ and $\PV^{(1)}(x)$ in Eq.~(2.8) are known for a long 
time \cite{FP82}, see also ref.~\cite{HvN} for the solution of an 
earlier problem in the covariant-gauge calculation of $P_{gg}^{(1)}$.  
The three-loop quantities $P_{ij}^{(2)}(x)$ are the subject of Sect.~4. 
The consistency of the evolution equations (2.3) with the momentum sum 
rule (2.2) imposes the following constraints on the second moments of 
$P_{ij}^{(l)}(x)\,$:
\bea
  P_{qq}^{(l)}(N) + P_{gq}^{(l)}(N)  & = & 0 \nonumber \\[1mm]
  P_{qg}^{(l)}(N) + P_{gg}^{(l)}(N)  & = & 0 \quad \mbox{for} \quad
  N = 2 \:\: ,
\eea
where
\beq
    a(N) \:\equiv\: \int_0^1 \! dx \, x^{N-1} a(x) \:\: .
\eeq
The constants $\beta_i$ in Eq.~(2.8) represent the perturbative 
coefficients of the QCD $\beta$-function 
\beq
  \frac{d a_s}{d \ln \mu_r^2} \: = \: \beta(a_s) \: = \:
  - \sum_{l=0} a_s^{l+2} \beta_l \:\: . 
\eeq
The coefficients $\beta_0,\, \beta_1$ and $\beta_2$ required for NNLO 
calculations can be found in refs.~\cite{FP82} and \cite{beta2}, 
respectively.
Finally the $L_R$ terms in Eq.~(2.8) are obtained from the expression 
for $\mu_r=\mu_f$ by inserting the expansion of $a_s(\mu_f^2)$ in terms 
of $a_s(\mu_r^2)$, 
\beq
  a_s(\mu_f^2) = a_s(\mu_r^2) - \beta_0 L_R\, a_s^2(\mu_r^2) 
  - \left\{ \beta_1 L_R - \beta_0^2 L_R^2 \right\} a_s^3(\mu_r^2) 
  + \ldots \:\: .
\eeq
  
The singlet structure functions $F_{a,S}$, $a = 1,\, 2$, are in 
Bjorken-$x$ space obtained by the convolution (2.4) of the solution
of Eq.~(2.3) with the corresponding coefficient functions,
\bea
  \eta_a F_{a,\rm S}(x,Q^2) & \! =\! & 
  \left[ {\cal C}_{a,q}(a_s, L_M, L_R) \otimes \Sigma(\mu_f^2, \mu_r^2) 
       + {\cal C}_{a,g}(a_s, L_M, L_R) \otimes g(\mu_f^2, \mu_r^2) 
  \right] (x) \quad \nonumber \\[1mm]
  & \!\equiv\! &  \left[ \CV_a(a_s, L_M, L_R) \otimes 
  \qV(\mu_f^2, \mu_r^2) \right] (x) \:\: .
\eea
Here the electroweak charge factor is included in $\eta_a$, e.g., 
$\eta_1 = 2\,\langle e^2 \rangle^{-1}$ and $\eta_2 = (x \,\langle 
e^2 \rangle)^{-1}$ for electromagnetic scattering, with an average
squared charge $\langle e^2 \rangle = 5/18 $ for an even $N_f$.  Up to 
third order in $\alpha_s$ the expansion of the coefficient functions 
takes the form 
\bea
  \CV_a (x, \alpha_s(\mu_r^2), L_M, L_R ) & = &
  \CV_a^{(0)}(x) \, +\, a_s \CV_a^{(1)}(x,L_M) \, +
  \nonumber \\ & & \mbox{}
  a_s^2 \CV_a^{(2)}(x,L_M,L_R) \, +\, a_s^3 \CV_a^{(3)}(x,L_M,L_R) 
  \, + \,\ldots \quad \\[1mm]
  & \!\!\stackrel{{\mu_r = \mu_f}}{=}\!\! & 
  \cV_a^{(0)}(x) \, + \,  \sum_{l=1}^{3} a_s^l \left( \cV_a^{(l)}(x) + 
  \sum_{m=1}^{l} \cV_a^{(l,m)}(x) L_M^m \right) \, + \,\ldots \:\: ,
  \quad \nonumber 
\eea
where $ \cV_a^{(0)}(x) \equiv (\, c_{a,q}^{(0)}(x)\, , \,  
c_{a,g}^{(0)}(x)\, ) = ( \,\delta (1-x)\, ,\, 0\, )$ represents the 
parton-model result. The first-order corrections $\cV_a^{(1)}(x)$ can 
be found in ref.~\cite{FP82}; the two-loop coefficient functions 
$c_{a,q}^{(2)}(x)$ and $c_{a,g}^{(2)}(x)$ computed in refs.~\cite
{ZvN1,ZvN2} are briefly addressed in Sect.~3.
The terms up to order $\alpha_s^2$ in Eq.~(2.16) contribute to $F_a$
in the NNLO approximation.  The $\alpha_s^3$ terms only enter for $dF_a 
/ d\ln Q^2$, hence the scale-independent three-loop quantities 
$\cV_a^{(3)}(x)$ are not needed here.
The coefficients $\cV_a^{(l,m)}(x)$ in the second part of Eq.~(2.16)
can be conveniently written in a recursive manner as
\bea
  \cV_a^{(1,1)} &\! =\! & \cV_a^{(0)} \otimes \PV^{(0)} \nonumber \\
  \cV_a^{(2,1)} &\! =\! & \cV_a^{(0)} \otimes \PV^{(1)} 
  + \cV_a^{(1)} \otimes (\PV^{(0)} - \beta_0 \eV) \nonumber \\
  \cV_a^{(2,2)} &\! =\! & \frac{1}{2} \cV_a^{(1,1)} \otimes 
  (\PV^{(0)} - \beta_0 \eV) \nonumber \\
  \cV_a^{(3,1)} &\! =\! & \cV_a^{(0)} \otimes \PV^{(2)} + \cV_a^{(1)} 
  \otimes (\PV^{(1)} - \beta_1 \eV) + \cV_a^{(2)} \otimes  
  (\PV^{(0)} - 2 \beta_0 \eV) \nonumber \\
  \cV_a^{(3,2)} &\! =\! & \frac{1}{2} \bigg\{ \cV_a^{(1,1)} \otimes 
  (\PV^{(1)} - \beta_1 \eV) + \cV_a^{(2,1)} \otimes (\PV^{(0)} - 
  2 \beta_0 \eV) \bigg\} \nonumber \\
  \cV_a^{(3,3)} &\! =\! & \frac{1}{3} \cV_a^{(2,2)} \otimes 
  (\PV^{(0)} - 2 \beta_0 \eV) \:\: .
\eea
Eqs.~(2.17) can be derived by identifying the results of the 
following two calculations of $dF_{a,S}/d\ln Q^2 $ at $\mu_r^2 =\mu_f^2 
= Q^2$: {\bf (a)} with the scales identified in the beginning (using 
Eqs.~(2.3) and (2.13), and {\bf (b)} with the scales set equal only at 
the end (after differentiating the logarithms in Eq.~(2.16)$\,$). 
Finally the coefficients $\CV_a^{(2)}$ and $\CV_a^{(3)}$ for 
$\mu_f \neq \mu_r$ in Eq.~(2.16) are obtained from these results by 
Eq.~(2.14):
\bea
  \CV_a^{(2)}(x,L_M,L_R) & \! =\! & \CV_a^{(2)}(x,L_M,0) - \beta_0 L_r  
  \,\CV_a^{(1)}(x,L_M) \nonumber \\[1mm]
  \CV_a^{(3)}(x,L_M,L_R) & \! =\! & \CV_a^{(3)}(x,L_M,0) 
  - 2 \beta_0 L_r \,\CV_a^{(2)}(x,L_M,0) 
  \\ & & \mbox{} 
  - \left\{ \beta_1 L_R - 
  \beta_0^2 L_R^2 \right\} \CV_a^{(1)}(x,L_M) \:\: . \nonumber 
\eea

The above-mentioned calculations of $\beta_2$ \cite{beta2} and 
$\cV_a^{(2)}(x)$ \cite{ZvN1,ZvN2} have been performed in the \MSb\ 
renormalization and factorization schemes. The other factorization 
scheme widely used in NLO analyses of unpolarized deep-inelastic 
scattering is the so-called DIS scheme. In this scheme the expression 
for $F_{2,S}$ is reduced, for the choice $\mu_r^2 = \mu_f^2 = Q^2$ 
adopted for the rest of this section, to its parton-model form 
$\eta_{2\,} F_{2,S}(x,Q^2) = \widetilde{\Sigma}(x,Q^2)$, i.e.,    
\beq
 \widetilde{c}_{2,q}^{\, (l)}(x) \:\equiv\: 
 \widetilde{c}_{2,g}^{\, (l)}(x) \: =\: 0 \quad \mbox{for} \quad  
 l \geq 1 \:\: .
\eeq 
Here and in the following DIS-scheme quantities are marked by a tilde. 
The corresponding singlet parton densities can be defined by
\beq
  \widetilde{\qV} \:\equiv\: \left( \begin{array}{c} 
  \!\widetilde{\Sigma}\! \\ \widetilde{g} \end{array} \right)
  \: = \: \left( \begin{array}{rr} c_{2,q} & c_{2,g} \\
  -c_{2,q} & -c_{2,g} \end{array} \right) \otimes
  \left( \begin{array}{c} \!\Sigma\! \\ g  \end{array} \right)
  \:\equiv\: \ZV \otimes \qV \:\: .
\eeq
The upper row of the transformation matrix $\ZV$ is fixed by 
Eq.~(2.19).  The lower row is constrained only by the momentum sum rule 
(2.2) which fixes its form for the second moment, $N = 2$. The 
extension of this constraint to all $N$, 
\beq
  \tilde{\Sigma}(N) + \tilde{g}(N) \: = \: \Sigma(N) + g(N) \:\: ,
\eeq
in Eq.~(2.20) is the obvious generalization of the standard NLO 
convention of ref.~\cite{DFLM} to all orders. The DIS-scheme splitting 
functions for $\mu_r^2 = \mu_f^2 = Q^2$ can be expressed in terms of 
their \MSb\ counterparts by differentiating of Eq.~(2.20) with respect 
to $Q^2$:
\beq
  \widetilde{\PV} \: =\:  \left( \ZV \otimes \PV + \beta \,
  \frac{d\ZV}{da_s} \right) \otimes \ZV^{-1} \:\: .
\eeq
Insertion of the expansions (2.8) for $L_R = 0$, (2.13) and
\beq
  \ZV(x,a_s) \: = \: \eV + \sum_{l=1} a_s^l \ZV^{(l)}(x) \: = \: 
  \eV + \sum_{l=1} a_s^l \left( \begin{array}{rr} c_{2,q}^{(l)}(x) 
  & c_{2,g}^{(l)}(x) \\[1mm] -c_{2,q}^{(l)}(x) & -c_{2,g}^{(l)}(x) 
  \end{array} \right) 
\eeq
then yields
\bea
  \widetilde{\PV} &\! =\! & \quad\: a_s \PV_0 
  \nonumber \\ & & \mbox{}
  + a_s^2 \left( \PV^{(1)} + [\ZV^{(1)},\PV^{(0)}] - \beta_0\ZV^{(1)} 
  \right) 
  \\ & & \mbox{} 
  + a_s^3 \left( \PV^{(2)} + [\ZV^{(2)},\PV^{(0)}] 
  + [\ZV^{(1)},\PV^{(1)}] - [\ZV^{(1)},\PV^{(0)}] \otimes \ZV^{(1)} 
  \right. \nonumber \\ & & \mbox{} \quad\quad
  + \left. \beta_0 \ZV^{(1)} \otimes \ZV^{(1)}
  - 2\beta_0 \ZV^{(2)} - \beta_1 \ZV^{(1)} \right) 
  \: + \: \ldots \:\: ,
  \nonumber
\eea
where
\beq
  [\ZV^{(l)},\PV^{(m)}] \equiv \ZV^{(l)} \otimes \PV^{(m)} 
  - \PV^{(m)} \otimes \ZV^{(l)} \:\: .
\eeq
The coefficient functions and splitting functions for $\mu_r^2 \neq 
\mu_f^2 \neq Q^2$ can by obtained from Eqs.~(2.19) and (2.24) using
Eqs.~(2.16)--(2.18) and (2.8), respectively. In the present article 
Eq.~(2.24) will be only employed in Sect.~4 for transforming a 
small-$x$ constraint obtained for $\widetilde{P}_{gg}^{(2)}$ 
\cite{FL98} into the \MSb\ scheme adopted for the rest of this paper. 
%
%
\section{The 2-loop singlet coefficient functions}
\setcounter{equation}{0}
%
%
Beyond the first order, $l=1$, the exact expressions for the 
scale independent parts $\cV_{a}^{(l)}(x)$ of the coefficient functions 
(2.16) are rather lengthy and involve higher transcendental 
functions.\footnote
{A recent recalculation of all these functions \cite{MV1} has yielded
 full agreement with refs.~\cite{ZvN1,ZvN2}.}
Moreover, the convolutions entering the expressions (2.17) for 
$\mu_f^2 \neq Q^2$ (given explicitly for $l=2$ in ref.~\cite{ZvN2}) and 
the factorization scheme transformation (2.24) become increasingly 
cumbersome at higher orders. The latter complications can be 
circumvented by using the moment-space technique \cite{DFLM,cmom}, as 
the convolution (2.4) reduces to a simple product in $N$-space. This 
technique, however, requires the analytic continuation of all 
ingredients to complex values of $N$ in Eq.~(2.12), which itself 
becomes rather involved already for the exact expressions for 
$\cV_{a}^{(2)}(x)$ \cite{BlKu}. Hence it is convenient, for both 
$x$-space and $N$-space applications, to dispose of compact 
parametrizations of these quantities.

For the non-singlet components $c_{L,\rm NS}^{(2)}(x)$ \cite{c2Sa} 
and $c_{2,\rm NS}^{(2)}(x)$ \cite{ZvN1} of 
\beq
  c_{a,q}^{(2)}(x) \: = \: c_{a,{\rm NS}}^{(2)}(x) + 
  c_{a,{\rm PS}}^{(2)}(x) \:\: , 
\eeq
where $F_L \equiv F_2 - 2xF_1$, such parametrizations have been given
in ref.~\cite{NV1} in terms of 
\beq
  L_0 \:\equiv\: \ln x \:\: , \quad L_1 \:\equiv\: \ln (1-x)
\eeq
and simpler functions of $x$. The pure singlet pieces in Eq.~(3.1) 
can be approximated by
\bea
  c_{L,{\rm PS}}^{(2)}(x)
  &=& N_f \:\Big\{ (15.94 - 5.212\,x) (1-x)^2 L_1 + 
      (0.421 + 1.520\, x) L_0^2 
      \nonumber \\
  & & \mbox{}+ 28.09\, (1-x) L_0 - (2.370\, x^{-1}-19.27) 
      (1-x)^3 \Big \} \:\: ,
  \\[2mm]
  c_{2,{\rm PS}}^{(2)}(x)
  &=& N_f \:\Big\{-0.101\, (1-x) L_1^3 - (24.75 - 13.80\, x) L_1 L_0^2
      + 30.23 L_1 L_0
      \nonumber \\
  & & \mbox{}+ 4.310\, L_0^3 - 2.086\, L_0^2 + 39.78\, L_0
      + 5.290 (x^{-1}-1) \Big\} \:\: .
\eea
The corresponding gluon coefficient functions \cite{ZvN1} can be 
parametrized as
\begin{eqnarray}
  c_{L,g}^{(2)}(x)
  &=& N_f \:\Big\{ (94.74 - 49.20\,x) (1-x) L_1^2 + 864.8\, (1-x) L_1 
      + 1161\, x L_1 L_0 
      \nonumber \\
  & & \mbox{}+ 60.06\, x L_0^2 + 39.66\, (1-x) L_0 
      - 5.333\, (x^{-1} - 1) \Big \} \:\: ,
  \\[2mm]
  c_{2,g}^{(2)}(x)
  &=& N_f \:\Big\{ (6.445 + 209.4\, (1-x) ) L_1^3
      - 24.00\, L_1^2 + (1494\, x^{-1} - 1483) L_1
      \nonumber \\
  & & \mbox{}+ L_1 L_0 (-871.8\, L_1 - 724.1 L_0)
      + 5.319 L_0^3 - 59.48\, L_0^2 - 284.8\, L_0
      \nonumber \\
  & & \mbox{} + 11.90\, x^{-1} + 392.4 - 0.28\, \delta (1-x) \Big\} 
      \:\: .
\end{eqnarray}
Note the small $\delta (1-x)$ term in the parametrization of 
$c_{2,g}^{(2)}$, which is of course absent in the exact expression, 
but useful for obtaining high-accuracy convolutions here.  The 
parametrizations (3.3)--(3.6) deviate from the exact results by no 
more than a few permille. 
%
%
\section{The 3-loop singlet splitting functions}
\setcounter{equation}{0}
%
%
Only partial results have been obtained so far for the $O(\alpha_s^3)$
terms $P_{ij}^{(2)}(x)$ of the singlet splitting functions (2.8).
As the full $x$-dependence is known just for the $C_A N_f^2$ 
part of $P_{gg}^{(2)}$ \cite{Gra2}, the backbone of the available 
constraints are the lowest four even-integer moments (2.12) of these 
terms,
\beq
  P_{ij}^{(2)}(N) \quad \mbox{for} \quad N \: =\:  2,\, 4,\, 6,\, 8
  \:\: ,
\eeq
calculated in ref.~\cite{spfm2}.  This is one moment less than 
determined for the non-singlet quantity $P_{\rm NS}^{(2)+}(x)$ 
\cite{spfm1,spfm2}, due to the greater technical complexity of the 
singlet computation.
 
The $1/[1-x]_+$ soft-gluon contributions to the gluon-gluon splitting 
functions $P_{gg}^{(0)}$ and $P_{gg}^{(1)}$ are related to their 
quark-quark counterparts by
\beq
  (1-x) P_{gg}^{(l)}(x)\bigg|_{x=1} 
  \: =\: \frac{C_A}{C_F}\: (1-x) P_{qq}^{(l)}(x)\bigg|_{x=1}
  \: =\: \frac{C_A}{C_F}\: (1-x) P_{\rm NS}^{(l)}(x)\bigg|_{x=1}  
\eeq
(${\cal P}_{\rm PS}^{\,}$ in Eq.~(2.5) vanishes at $x = 1$).  This 
relation also applies to the leading-$N_f$ terms \cite{Gra1,Gra2} of 
$P_{gg}^{(2)}$ and $P_{\rm NS}^{(2)}$.  We will assume that Eq.~(4.2) 
holds generally for $l = 2$.

The small-$x$ behaviour of $P_{ij}^{(2)}(x)$ is not constrained by 
Eq.~(4.1). The leading $x \!\rightarrow\! 0$ terms $\propto \frac{1}{x}
\ln x$ have been determined from small-$x$ resummations, however, for 
$P_{qq}^{(2)}$, $P_{qg}^{(2)}$ and $P_{gg}^{(2)}$.  Specifically, the 
gluon-quark and quark-quark entries read \cite{CH94}
\bea
  P_{qg, x \rightarrow 0}^{(2)} & = & \mbox{} \!\! - \frac{896}{27}\:
  C_A^2 N_f \:\frac{\ln x}{x} \: + \: O \left( \frac{1}{x} \right) 
  \:\: , \\
  P_{qq, x \rightarrow 0}^{(2)} & = & \quad \frac{C_F}{C_A}\:
  P_{qg, x \rightarrow 0}^{(2)} \:\: + \: O \left( \frac{1}{x} \right) 
  \:\: .
\eea 
The corresponding gluon-gluon result can be inferred \cite{BNRV} from 
the larger eigenvalue of $\PV_{x \rightarrow 0}^{(2)}$ completed in 
ref.~\cite{FL98} in a scheme equivalent to the DIS scheme up to 3-loop 
order: 
\beq
  \widetilde{P}_{gg, x \rightarrow 0}^{(2)} = \left\{ 6320 - 
  864\,\zeta(3) - 1584\,\zeta(2) + N_f \left( \frac{1136}{3} 
  - 96\, \zeta(2) \right) \right\} \frac{\ln x}{x} \: + \: 
  O \left( \frac{1}{x} \right) \:\: ,
\eeq
where $\zeta(2) = \pi^2/6$ and $\zeta(3) \simeq 1.202057$. For this 
contribution the scheme transformation (2.24) schemes simplifies to 
\beq
  \widetilde{P}_{gg,x\rightarrow 0}^{(2)} \: = \: 
  P_{gg,x\rightarrow 0}^{(2)} - \Big( P_{gq}^{(0)}\otimes c_{2,g}^{(2)} 
  \Big)_{x\rightarrow 0} \: + \: O \left( \frac{1}{x} \right) \:\: ,
\eeq
yielding
\beq
  P_{gg, x \rightarrow 0}^{(2)} \: = \: \left\{ 6320 - 864\,\zeta(3) - 
  1584\,\zeta(2) + N_f \left( \frac{4720}{27} - \frac{32}{3}\, \zeta(2) 
  \right) \right\} \frac{\ln x}{x} \: + \: O \left( \frac{1}{x} \right)
  \:\: . 
\eeq
%
Eqs.~(4.6) and (4.7) are most easily derived via moment-space, see the
appendix.

In the following the above information will be used for approximate
representations of 
\beq
  P_{ij}^{(2)}(x) \: =\:  P_{ij,0}^{(2)}(x) + N_f P_{ij,1}^{(2)}(x)
  + N_f^2 P_{ij,2}^{(2)}(x) \:\: ,
\eeq
where the $N_f$-independent terms are absent in $P_{\rm PS}^{(2)}$ 
and $P_{qg}^{(2)}$ according to Eqs.~(2.6) and (2.7). For each of the
ten remaining contributions $P_{ij,m}^{(2)}$ we employ the ansatz
\beq
   P_{ij,m}^{(2)}(x) \: =\: \sum_{n=1}^{4} A_n f_n(x) + f_e(x) \:\: .
\eeq
Except for the $\delta (1-x)$ terms in $P_{gg,m}^{(2)}$, the basis 
functions $f_n$ are build up of powers of $\ln (1-x)$, $x$, and 
$\ln x$. Generally we choose the functions $f_n$ to include at least 
one function peaking as  $x \ra 1$ (except for $P_{\rm PS}^{(2)}$, of
course), one rather flat function (a small non-negative power of $x$), 
and one function peaking as $x \ra 0$. For each choice of the building
blocks $f_n$, the coefficients $A_n$ are determined from the four 
linear equations provided by the moments (4.1) of ref.~\cite{spfm2} 
after taking into account the additional information (4.2)--(4.4) and 
(4.7) collected in the function $f_e$ in Eq.~(4.9).
The residual uncertainty of the functions $P_{ij,m}^{(2)}$ is estimated 
by varying the choice of the basis functions $f_n$. Finally two 
approximate expressions are selected for each of the six quantities 
$P_{gq,0\,}^{(2)}$, $P_{gg,0}^{(2)}$ and $P_{ij,1}^{(2)}$ which are 
representative for the respective present uncertainty band.  The 
$N_f^2$ pieces $P_{ij,2}^{(2)}$ are smaller in absolute size and 
uncertainty than those contributions, hence for them it suffices to 
select just one central representative.

Before proceeding to our approximations of the coefficients 
$P_{ij,m}^{(2)}$ in Eq.~(4.8), it is worthwhile to discuss another 
constraint on the singlet splitting functions which has served as an 
important check for the calculations of the NLO quantities 
$P_{ij}^{(1)}(x)$ in the unpolarized \cite{FP82} as well as in the 
polarized case \cite{MvN,WV96}. This is the relation
\beq
  P_{qq,{\rm DR}}^{(l)}(x) + P_{gq,{\rm DR}}^{(l)}(x) - 
  P_{qg,{\rm DR}}^{(l)}(x) - P_{gg,{\rm DR}}^{(l)}(x) \: = \: 0
\eeq
for the choice
\beq
  C_A \: \equiv \: N_c \: =\: C_F \: =\: N_f \nonumber
\eeq
of the colour factors leading to a ${\cal N}\! =\! 1$ supersymmetric 
theory.  As indicated, Eq.~(4.10) holds in the dimensional reduction 
(DR) scheme respecting the supersymmetry, but not in the usual \MSb\ 
scheme based on dimensional regularization. The terms breaking the 
relation (4.10) in the \MSb\ scheme for $l=1$ are, however, much 
simpler than the functions $P_{ij}^{(1)}(x)$ themselves. They involve 
only $\ln^1 x$ and powers of $x$, however including the leading \mbox
{small-$x$} term $1/x$ \cite{FP80}. It should also be noted that 
Eq.~(4.10) does not in all cases introduce a relation between different 
functions $P_{ij}^{(l)}$. For instance, the leading large-$x$ terms of 
$P_{gq}^{(1)}$ and $P_{qg}^{(1)}$, $\ln^2 (1-x)$ and $\ln (1-x)$,  
cancel already on the level in the individual splitting functions for 
the choice (4.11), not only after forming the combination (4.10), as 
can be readily read off from the tables in refs.~\cite{CFP,EV96}.

The transformation of the splitting functions from the DR\footnote
{Actually the construction of a consistent DR scheme respecting the
 supersymmetry at 3-loop level has not been performed so far 
 \cite{AV83}.  Consequently Eq.~(4.10) has not been proven up to now 
 for $l\! >\! 1$.}
to the \MSb\ scheme is presently unknown for $l=2$ (its derivation 
would require repeating the calculation of the finite terms of the 
two-loop operator matrix elements of ref.~\cite{MSvN} in the DR 
scheme).  It is expected, however, that the terms breaking Eq.~(4.10) 
in the \MSb\ scheme at $l=2$ are again comparatively simple and do not 
include terms like $\ln^3 (1-x)$, $\ln^4 (1-x)$, and $\ln^4 x$. Under 
this assumption that relation does provide some constraints on the 
functions $P_{ij}^{(2)}$.  In view of their limitations discussed above
for the NLO case, however, these few additional constraints do not 
justify moving from Eq.~(4.8) to a complete colour-factor decomposition 
involving twenty-three instead of ten unknown functions 
$P_{ij,m}^{(2)}$. 

We now illustrate the details of our approximation procedure outlined
above for the case of the gluon-quark splitting functions $P_{qg}$ 
dominating the small-$x$ evolution of the singlet quark density.  While 
the lowest-order result $P_{qg}^{(0)}$ contains no logarithms, terms up 
to $\ln^2 (1-x)$ and $\ln^2 x$ occur in $P_{qg}^{(1)}$. Hence we expect 
contributions up to $\ln^4 (1-x)$ and $\ln^4 x$ in the three-loop 
splitting function $P_{qg}^{(2)}$ due to the additional loop or 
emission.  Thus a reasonable choice of the trial functions
for $P_{qg,1}^{(2)}$ is given by
\beq
  \begin{array}[b]{ccccc}
  f_1(x) &=& \ln^2 x     & \mbox{ or } & \ln^3 x     \\[0.5mm]
  f_2(x) &=&   1         & \mbox{ or } &    x        \\[0.5mm]
  f_3(x) &=& \ln\, (1-x) & \mbox{ or } & \ln^2 (1-x) \\[0.5mm]
  f_4(x) &=& \ln^3 (1-x) & \mbox{ or } & \ln^4 (1-x) \\[2mm]
  f_e(x) &=& \multicolumn{3}{l} { {\displaystyle -
             \frac{896}{27}\, C_A^2 N_f \left( \frac{\ln x}{x} 
             + \frac{\lambda}{x} \right) } }
  \end{array} , \quad \lambda = 0\:\mbox{ or }\: 4 \:\: .
\eeq
As for all other cases in which the $\frac{1}{x}\ln x$ terms are known,
we do not include the subleading contribution $1/x$ in the functions
$f_n$, but vary its coefficient by hand up to a value suggested by the 
expansion of the moment-space expressions $P_{ij}^{(0)}(N)$ and 
$P_{ij}^{(0)}(N)$ around $N=1$ \cite{BRV}. Four of the thirty-two
combinations resulting from Eq.~(4.12) (those combining $f_1 = \ln^2 x$ 
with $f_2 = 1$ and $\lambda = 4$) lead to an almost singular system for 
the coefficients $A_n$ which causes overly large oscillations of the 
corresponding results for $P_{qg,1}^{(2)}(x)$. The remaining 
twenty-eight approximations, of which four are selected for further 
consideration, are displayed in Fig.~1. 

\begin{figure}[ht]
\vspace*{1mm}
\centerline{\epsfig{file=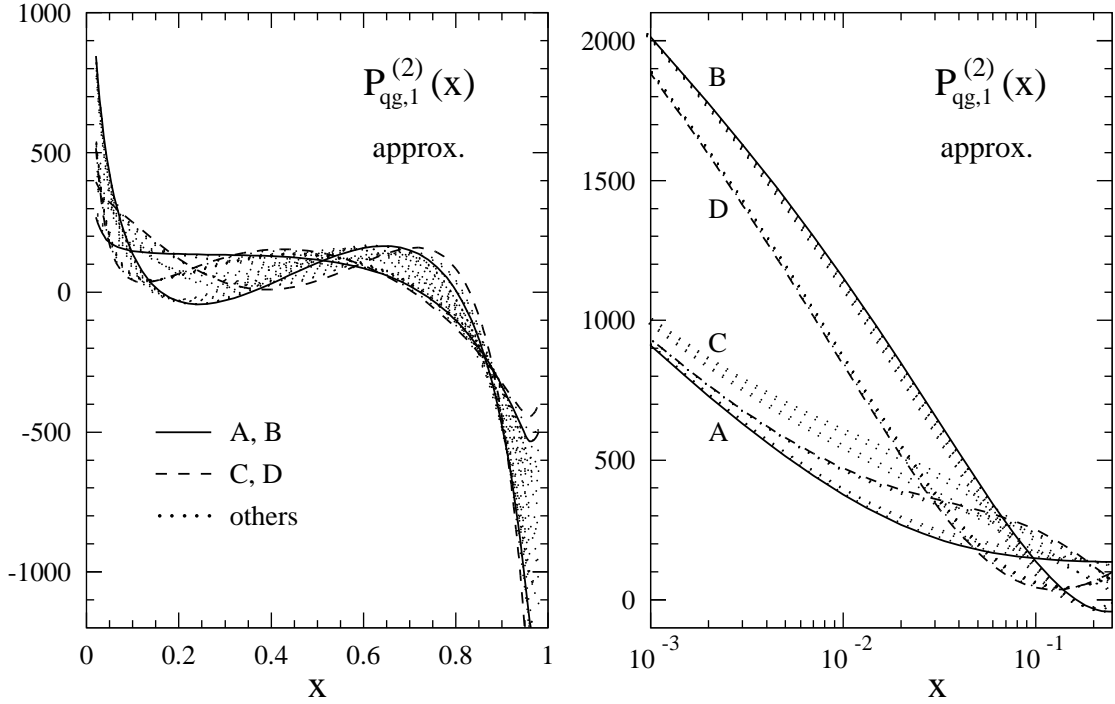,width=15cm,angle=0}}
\vspace{-1.5mm}
\caption{Approximations of the $N_f^1$ part $P_{qg,1}^{(2)}$ of the
 three-loop splitting function $P_{qg}^{(2)}(x)$ in Eq.~(2.8), as 
 obtained from the four moments (4.1) by means of Eqs.~(4.9) and 
 (4.12).  The full and dashed curves represent the functions selected 
 for further consideration. The upper group of curves in the right plot
 is for $\lambda=0$ in Eq.(4.12), the lower group for $\lambda = 4$.}
\vspace{1mm}
\end{figure}

\begin{figure}[ht]
\vspace{1mm}
\centerline{\epsfig{file=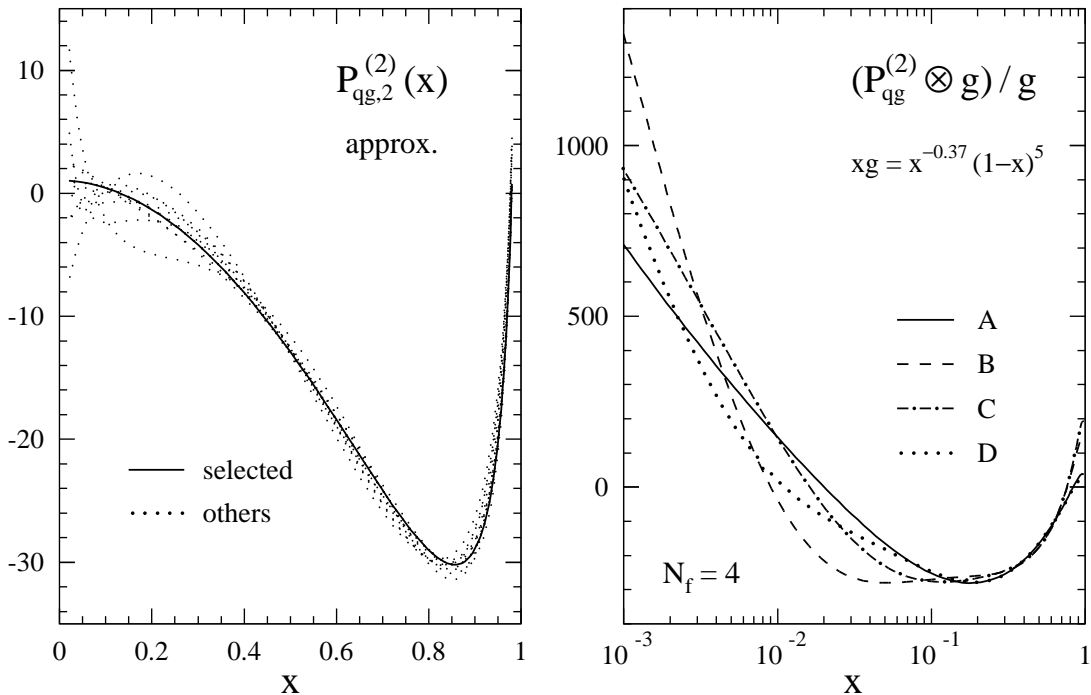,width=15cm,angle=0}}
\vspace{-1.5mm}
\caption{Left: as Fig.~1, but for the $N_f^2$ contribution 
 $P_{qg,2}^{(2)}(x)$ using  Eqs.~(4.9) and (4.13). 
 Right: convolution of the approximations A--D of Fig.~1, combined 
 for $N_f = 4$ with the result selected for $P_{qg,2}^{(2)}$ in the 
 left part, with a typical gluon density of the proton.} 
\end{figure}

Our corresponding ansatz for the $N_f^2$ part reads
\beq
  \begin{array}[b]{ccccccc}
  f_1(x) &=& \ln x       & \mbox{ or } & \ln^2 x     & \mbox{ or } & 
             \ln^3 x  \\ [0.5mm]
  f_2(x) &=&   1         & \mbox{ or } &    x        & & \\[0.5mm]
  f_3(x) &=& \ln (1-x)   & \mbox{ or } & \ln^2 (1-x) & \mbox{ or } &  
             \ln^3 (1-x) \\[0.5mm]
  f_4(x) &=& {\displaystyle \frac{1}{x}} & \mbox{and} 
             &\multicolumn{2}{c}{f_e(x) \: = \: 0 \:\: .} & 
  \end{array} 
\eeq
Here and for all other $N_f^2$ terms the $1/x$ contribution is included 
in the functions $f_n$, as no information about its magnitude is 
available. Fifteen of the eighteen combinations resulting from 
Eq.~(4.13) are shown in the left part of Fig.~2. The three combinations 
involving $f_1 = \ln^3 x$ and $f_2 = x$ are rejected for the same 
reason as the four expressions for $P_{qg,1}^{(2)}$ mentioned above.

In the right part of Fig.~2 the approximations selected for 
$P_{qg,1}^{(2)}$ and $P_{qg,2}^{(2)}$ have been combined into Eq.~(4.8) 
for $N_f = 4$ and convoluted with a typical example for the gluon 
density of the proton.  Needless to say that it is this combination 
$P_{qg}^{(2)} \otimes g$ (divided by $g(x)$ for display purposes in 
Fig.~2) and not $P_{qg}^{(2)}(x)$ itself which enters the evolution
equations (2.3) and thus the structure functions (2.15). 
At moderately large $x$, $0.1\! <\! x\! <\! 0.7$, the spread of the 
convolutions in Fig.~2 is small, typically about 5\%, much smaller than 
that for the splitting function itself, as the convolution (2.4) tends 
to wash out the oscillating differences of Fig.~1 to a large extent. 

The large-$x$ behaviour of the splitting functions is very important 
also at small $x$, as can be inferred from a comparison of Figs.~1 
and~2: while $P_{qg}^{(2)}(x)$ is dominated at $x < 10^{-2}$ by the
$\frac{1}{x}\ln x$ and $1/x$ terms, this does by no means hold
for $[P_{qg}^{(2)}\otimes g](x)$. For instance, the leftmost cross-over 
of the approximations C ($\lambda\! =\! 4$) and D ($\lambda\! =\! 0$) 
for $P_{qg}^{(2)}$ is at $x = 0.03$, but the convolution for the 
function D does not exceed that for the approximation C above 
$x = 10^{-3}$.  An even more striking illustration can be obtained by 
keeping only the small-$x$ function $f_e$ in Eq.~(4.12) for $P_{ij,1}$. 
The convolution then yields $(P_{qg}^{(2)}\otimes g)/g \simeq 4000$ and 
$-6000$ for $\lambda = 0$ and $\lambda = 4$ at $x = 10^{-3}$, 
respectively, instead of the spread between about 700 and 1300 found 
after taking into account the large-$x$ constraints (4.1).

Except close to the unavoidable cross-over points, the functions A and 
B are representative for the uncertainty bands of Figs.~1 and~2. 
Hence these approximations are chosen as our final results for the 
gluon-quark splitting function:
\bea
  P^{(2)A}_{{qg},\, 1}(x)\! &=&
     - 19.5515\: L_1^3 + 707.438\: L_1 + 2300.986 + 814.928\: L_0^2
     - \frac{896}{3}\: \frac{L_0 + 4}{x} \quad\quad\quad 
  \nonumber \\[1mm]
  P^{(2)B}_{{qg},\, 1}(x)\! &=&
     10.8972\: L_1^4 - 315.331\: L_1^2 + 902.843\: x - 1054.09\: L_0^2
     - \frac{896}{3}\: \frac{L_0}{x} 
\eea
with $L_0$ and $L_1$ as defined in Eq.~(3.2). The selected 
approximation for the $N_f^2$ piece reads
\beq
  P^{(2)}_{{qg},\, 2}(x)\: =\: 
  3.769\: L_1^2 - 59.176\: x + 0.244\: L_0 + 1.079\: \frac{1}{x}  
  \:\: .
\eeq
As for all other cases listed below, the average $\frac{1}{2}[A+B]$
represents our central result.

\begin{figure}[bth]
\centerline{\epsfig{file=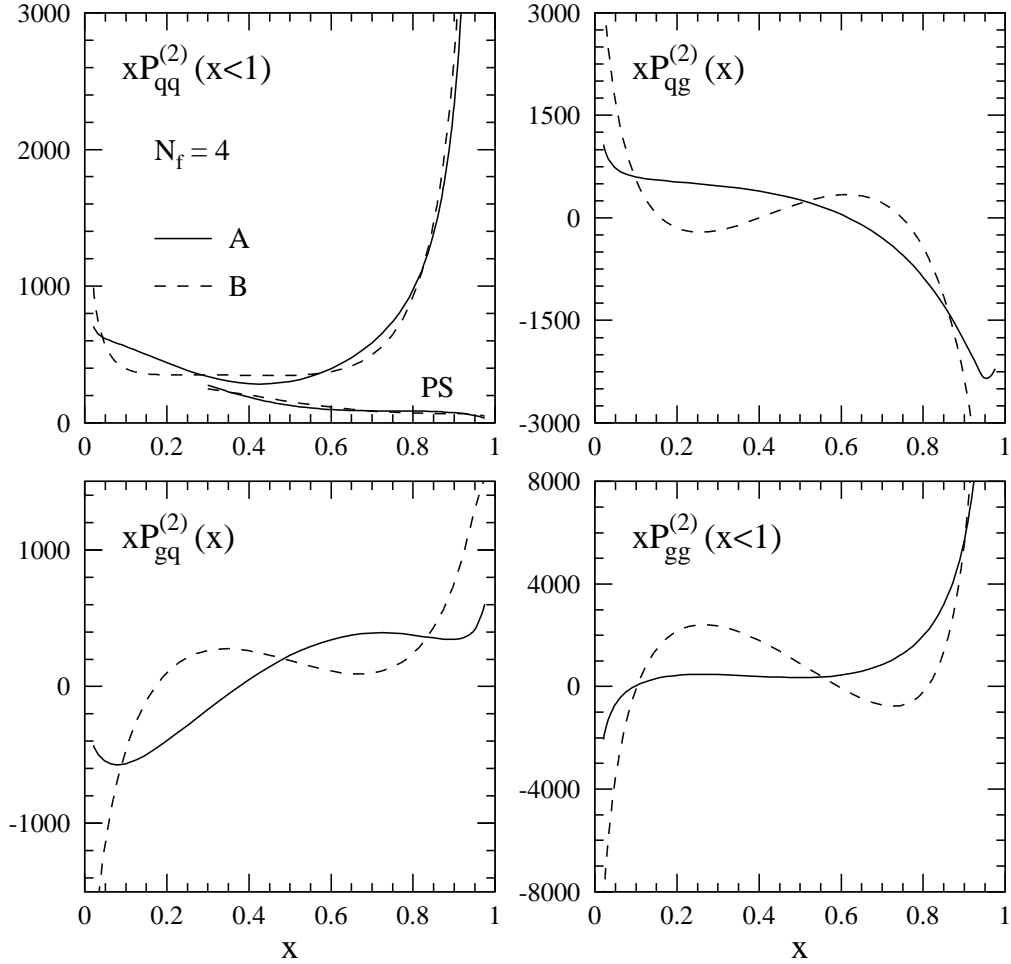,width=13.8cm,angle=0}}
\vspace{-2mm}
\caption{The large-$x$ behaviour of our selected approximations 
 of the three-loop splitting functions $P_{ij}^{(2)}(x)$ for $N_f=4$. 
 $P_{qq}^{(2)}$ is obtained by adding the pure singlet (PS) term
 (also shown separately) to the non-singlet contribution of ref.\ 
 \cite{NV1} according to Eq.~(2.5).}
\end{figure}

We now turn to the other splitting functions $P^{(2)}_{\rm PS}$, 
$P^{(2)}_{gg}$ and $P^{(2)}_{gq}$. Here we confine ourselves for 
brevity to the final results corresponding to Eqs.~(4.14) and (4.15) 
and some brief remarks on the individual cases. The resulting $N_f = 4$
approximations are graphically displayed in Fig.~3 for large $x$ and in 
Fig.~4 for moderately small $x$.
 
The approximations selected for the pure singlet splitting function 
are given by
\bea
  P^{(2)A}_{{\rm PS},\, 0}(x)\! &=&
     (1-x)\, ( -93.265\: L_1 + 357.924\: x ) + 543.482\: L_0^2
     + 9.864\: L_0^3 
  \nonumber \\ & & \mbox{}
     - \frac{3584}{27}\: \frac{1}{x} [ L_0 + 4\, (1-x) ] 
  \\
  P^{(2)B}_{{\rm PS},\, 1}(x)\! &=&
     (1-x)\, ( 37.395\: L_1^2 - 210.424\: x ) - 171.7\: L_0^2 
     - 48.862\: L_0^4 - \frac{3584}{27}\: \frac{L_0}{x} \quad\quad  
  \nonumber
\eea
and
\beq
  P^{(2)}_{{\rm PS},\, 2}(x)\: =\: 
     (1-x)\, ( - 3.999\: L_1 - 12.541\: x ) - 8.852\: L_0  
     + 3.445\: \frac{1}{x}(1-x) \:\: .
\eeq
The lowest-order pure-singlet contribution $P_{\rm PS}^{(1)}$
vanishes like $(1-x)^3$ as $x \ra 1 $.  The one additional loop or 
emission in $P^{(2)}_{\rm PS}$ will introduce logarithms up to $\ln^2 
(1-x)$, but keep $P_{\rm PS}^{\,}(x\! =\! 0) = 0$.  We have 
implemented this feature into all the basis functions $f_n$ in 
Eq.~(4.9), e.g., the large-$x$ logarithms are introduced as $(1-x) 
\ln^a (1-x)$.  The detailed $x \ra 1 $ behaviour of $P^{(2)}_{\rm PS}$ 
is however irrelevant in view of the large-$x$ dominance of the 
non-singlet contribution obvious from Fig.~3.

The corresponding results for the gluon-gluon splitting function read
\begin{eqnarray}
  P^{(2)A}_{{gg},\, 0}(x)\! &=&
     2560\: \frac{1}{(1-x)}_+ + 3870.26\: \delta (1-x) + 1292.56\: L_1
     \nonumber \\ & & \mbox{}
     - 14903.16 - 3667.22\: L_0^2 + 2675.85\: \frac{1}{x} (L_0 + 4)
     \\[1mm]
  P^{(2)B}_{{gg},\, 0}(x)\! &=&
     3031\: \frac{1}{(1-x)}_+ + 5622.22\: \delta (1-x) - 14514.35\: x^2
     \nonumber \\ & & \mbox{}
     + 643.44 + 13565.99\: L_0^2 + 2675.85\: \frac{L_0}{x} 
     \:\: , \nonumber 
\end{eqnarray}
\begin{eqnarray}
  P^{(2)A}_{{gg},\, 1}(x)\! &=&
     - 427.5\: \frac{1}{(1-x)}_+ - 570.4\: \delta (1-x) 
     + 2529.794\: x^2 
     \nonumber \\ & & \mbox{} 
     - 1605.009 - 784.828\: L_0^2 + 157.18\: \frac{1}{x} (L_0 + 4) 
     \\[1mm]
  P^{(2)B}_{{gg},\, 1}(x)\! &=&
     - 405.0\: \frac{1}{(1-x)}_+ - 539.16\: \delta (1-x) 
     - 929.17\: L_1  
     \nonumber \\ & & \mbox{} 
     - 1345.962\: x + 21.917\: L_0^2 + 157.18\: \frac{L_0}{x} \nonumber
\end{eqnarray}
and
\begin{equation}
  P^{(2)}_{{gg},\, 2}(x)\: =\: 
     - \frac{16}{9}\: \frac{1}{(1-x)}_+ + 6.575\: \delta (1-x) 
     - 5.056\: x - 9.904\: L_0 + 2.969\: \frac{1}{x} \:\: .
\end{equation}
The approximate $1/[1-x]_+$ coefficient in Eqs.~(4.18) and (4.19) have 
been taken over, according to Eq.~(4.2), from the results (4.9)--(4.11) 
for the non-singlet splitting functions in ref.~\cite{NV1}. The 
corresponding contribution to $P^{(2)}_{gg,2}$ is an exact 
leading-$N_f$ result of ref.~\cite{Gra2}, as the second colour factor 
of $P^{(2)}_{{gg},\, 2}$ not determined there does not contain a 
$1/[1-x]_+$ term. The $N_f^0$ and $N_f^1$ approximations have been 
combined in such a manner as to maximize the error band where the 
present uncertainties are largest, i.e., at small $x$.

\begin{figure}[thb]
\vspace*{-1mm}
\centerline{\epsfig{file=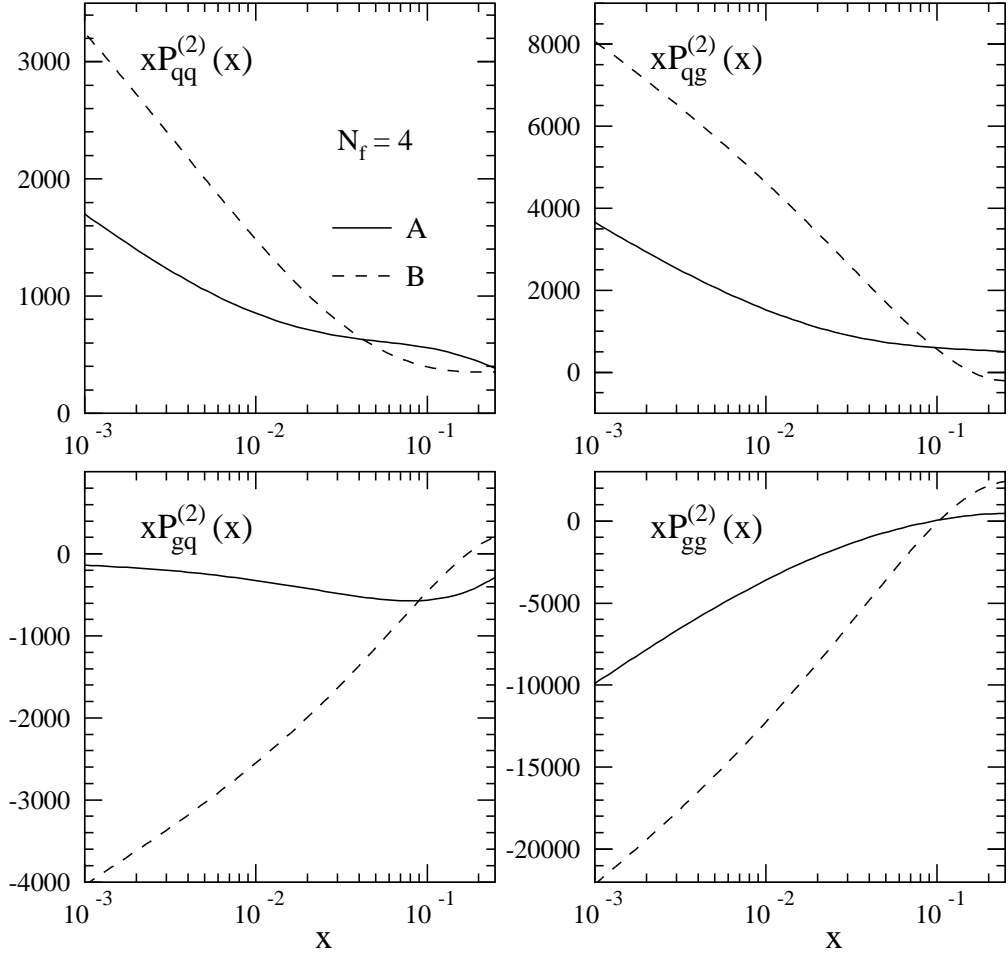,width=13.8cm,angle=0}}
\vspace{-2mm}
\caption{As Fig.~3, but for the small-$x$ behaviour of 
 $P_{ij}^{(2)}(x)$.  The difference between $P_{qq}^{(2)}$ and 
 $P_{\rm PS}^{(2)}$ is negligible for $x < 0.1$, hence the latter 
 quantity is not separately shown here.} 
\end{figure}

Finally the following expressions are chosen for the quark-gluon 
splitting function
\begin{eqnarray}
  P^{(2)A}_{{gq},\, 0}(x)\! &=&
     3.040\: L_1^3 + 1157.76 + 2357.73\: L_0^2 + 291.76\: \frac{L_0}{x}
     \nonumber \\
  P^{(2)B}_{{gq},\, 0}(x)\! &=&
     - 6.461\: L_1^3 - 1789.06\: x^2 + 2260.38 - 140.18\: \frac{L_0}{x}
     \:\: ,
\end{eqnarray}
\begin{eqnarray}
  P^{(2)A}_{{gq},\, 1}(x)\! &=&
     26.2717\: L_1^3 + 148.036\: L_1^2 - 549.815\: x + 89.769\: L_0^3 
     + 70\: \frac{L_0}{x} \nonumber \\
  P^{(2)B}_{{gq},\, 1}(x)\! &=&
     0.1995  \: L_1^3 - 37.93\: L_1 - 179.644 - 312.616\: L_0^2
     + 35\: \frac{L_0}{x}  
\end{eqnarray}
and
\begin{equation}
  P^{(2)}_{{gq},\, 2}(x)\: =\: 
  - 2.728\: L_1 - 10.217\: x - 3.566\: L_0 - 4.207\: \frac{1}{x} \:\: .
\end{equation}
Here the coefficient of leading small-$x$ term $\frac{1}{x}\ln x$ is  
unknown. The leading $x \ra 0$ contribution $\propto 1/x$ to the LO 
splitting function $P_{gq}^{(0)}$ is related to the corresponding term 
of $P_{gg}^{(0)}$ by a factor $C_F/C_A$. The same holds, up to two 
percent, for the $N_f$-parts of the NLO quantities $P_{gq}^{(1)}$ and 
$P_{gg}^{(1)}$ (but not for the $N_f^0$ parts). Therefore we have 
included $\frac{1} {x}\ln x$ in the functions $f_n$ in Eq.~(4.9) for 
$P^{(2)}_{gq,0}$, but varied its coefficient by hand for 
$P^{(2)}_{gq,1}$ between $ C_F/(2\,C_A)$ and $C_F/C_A$ relative to the 
corresponding term of $P^{(2)}_{gg,1}$ in Eq.~(4.19).  The resulting 
larger small-$x$ uncertainty of $P^{(2)}_{gq}$ is clearly visible in 
Fig.~4. While the uncertainty of the other splitting functions relative 
to the central results not shown in the figure does not exceed 
$\pm$30--40\% at $x = 10^{-3}$, it amounts to about 100\% for 
$P^{(2)}_{gq}$.
%
%
\section{Numerical results}
\setcounter{equation}{0}
%
%
We now proceed to illustrate the numerical effects of the NNLO 
contributions on the evolution of the singlet parton densities $\Sigma 
(x,\mu_f^2)$ and $g(x,\mu_f^2)$ and on the singlet structure function 
$F_{2,S}(x,Q^2)$.  Specifically, we will consider the scale derivatives
$\dot{q} \equiv d\ln q / d\ln \mu_f^2$, $q = \Sigma ,\, g$, at a fixed 
reference scale $\mu_f = \mu_{f,0\,}$, and $F_{2,S}$ and its 
$Q^2$-derivative at $Q^2 = \mu_{f,0}^2$.  Except for the last part of 
our discussions below, these illustrations will be given for the fixed, 
i.e.\ order-independent, initial distributions
\bea
 x\Sigma (x,\mu_{f,0}^2) &\! =\! & 0.6\, x^{-0.3}\, (1-x)^{3.5} 
                                   (1 + 5\, x^{0.8})  \:\: , 
 \nonumber \\
 xg (x,\mu_{f,0}^2) &\! =\! & 1.0\, x^{-0.37} (1-x)^{5} \:\: .
\eea
These expressions agree well with standard NLO parametrizations 
\cite{MRS,CTEQ,GRV}, at our choice $\mu_{f,0}^2 \simeq 30 
\mbox{ GeV}^2$. Likewise we employ, for the time being, 
\beq
 \alpha_s (\mu_r^2 = \mu_{f,0}^2) \: = \: 0.2
\eeq
irrespective of the order of the expansion, corresponding to $\alpha_s 
(M_Z^2) \simeq 0.116 $ beyond LO. All results below refer to the 
$\overline{\mbox{MS}}$ scheme for $N_f=4$ massless quark flavours. 
See ref.~\cite{CSvN} for a recent discussion of charm mass effects in 
$F_2$.

\begin{figure}[thbp]
\centerline{\epsfig{file=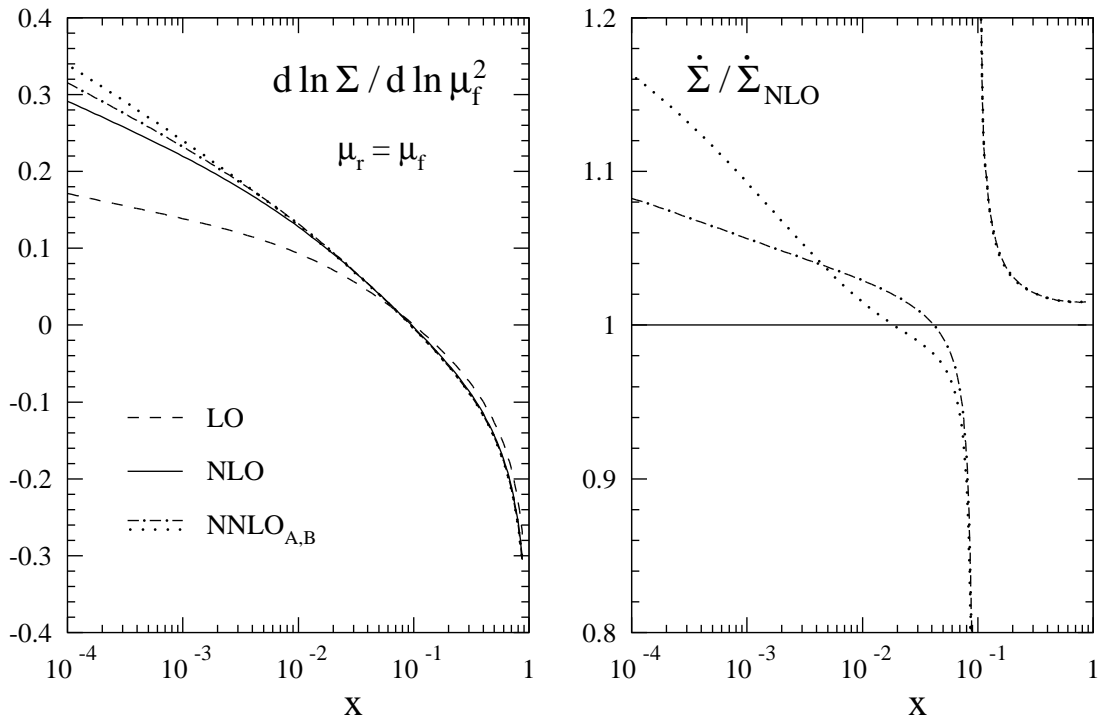,width=15cm,angle=0}}
\vspace{-2mm}
\caption{The perturbative expansion of the scale derivative, 
 $\dot{\Sigma} \equiv d \ln \Sigma / d\ln \mu_f^2$, of the singlet 
 quark density at $\mu_f^2 = \mu_{f,0}^2 \simeq 30 \mbox{ GeV}^2$. 
 The initial conditions are specified in Eqs.~(5.1) and (5.2).} 
\end{figure}

\begin{figure}[thbp]
\centerline{\epsfig{file=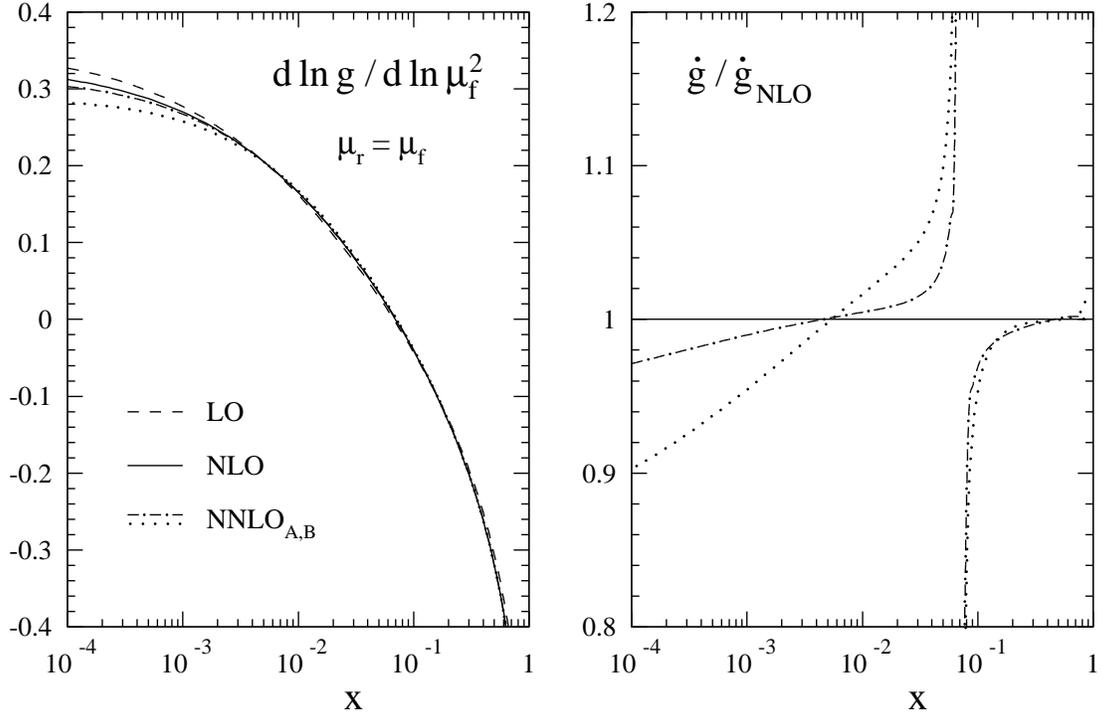,width=15cm,angle=0}}
\vspace{-2mm}
\caption{As Fig.~5, but for the gluon density $g(x,\mu_{f,0}^2)$. The 
 spikes close to $x = 0.1$ in the right parts of both figures
 are due to zeros of the NLO predictions.} 
\end{figure}

The LO, NLO and NNLO approximations to the evolution (2.3) of the 
singlet quark and gluon densities are compared in Fig.~5 and Fig.~6, 
respectively, for the standard choice $\mu_r = \mu_f$ of the 
renormalization scale. Here and in the following figures the subscripts 
A and B refer to the approximate expressions for the functions 
$P_{ij}^{(2)}(x)$ derived in the previous section; the central results 
$\frac{1}{2}(\mbox{NNLO}_A + \mbox{NNLO}_B)$ are not shown separately. 
For the input (5.1) and (5.2), the singlet quark derivative 
$\dot{\Sigma}$ is dominated by the $P_{qq} \otimes \Sigma$ contribution 
at large $x$, $x > 0.3$, and by radiation from gluons at small $x$, 
$x < 0.03$ (here the NLO corrections are very large, partly due to the 
absence of $1/x$ terms in $P_{qi}^{(0)\,}$ discussed above and below
Eq.~(2.7)$\,$). The gluon 
derivative $\dot{g}$, on the other hand, is mainly driven by the 
$P_{gg} \otimes g$ term, but the radiation from quarks is 
non-negligible over the full $x$-range. 

In both cases the NNLO corrections are small at large $x$, amounting to 
about 2\% for $\dot{\Sigma}$ and less than 1\% for $\dot{g}$. The 
corresponding NLO contributions are $12 - 20\%$ and $2 - 4\%$, 
respectively.  The present residual uncertainties of $P_{ij}^{(2)}$ 
are completely immaterial in this region of $x$. The NNLO effects and 
their uncertainties increase towards very small $x$-values, reaching 
about $(7.5\pm 2)\%$ for $\dot{\Sigma}$ and $(3 \pm 2)\%$ for $\dot{g}$ 
at $x= 10^{-3}$. Recall that these numbers refer to $\mu_{f,0}^2 \simeq 
30\mbox{ GeV}^2$. 
At lower scales the small-$x$ shapes of the quark and gluon densities 
are flatter than in Eq.~(5.1). Together with a larger $\alpha_s$ this 
leads to larger small-$x$ uncertainties.  For example, at $\mu_{f}^2 
\simeq 3\mbox{ GeV}^2$ (corresponding to $\alpha_s \simeq 0.3$) they 
reach about $\pm 6\%$ for $x= 10^{-3}$ and fall below $\pm 2\%$ only 
for $x\gsim 4\otimes 10^{-3}$. 

The renormalization scale dependence of the NLO and NNLO predictions 
for the derivatives $\dot{\Sigma}$ and $\dot{g}$ is shown in Fig.~7 and 
Fig.~8, respectively, for six representative values of~$x$. In both 
figures $\mu_r$ is varied over the rather wide interval $\frac{1}{8}\,
\mu_{f,0}^2 \leq\mu_r^2 \leq 8 \,\mu_{f,0}^2$ corresponding to $\, 0.29 
\,\gsim\, \alpha_s(\mu_r^2) \,\gsim\, 0.15\, $ for the input (5.2). 
The present approximate NNLO predictions prove sufficient for a marked 
improvement on the NLO results, except for the bins $x = 0.05$ (where 
the derivatives of both the singlet quark and gluon densities are 
small) and, in the gluon case, $x = 10^{-3}$. As illustrated above the 
$P_{ij}^{(2)}$ parametrization uncertain\-ties of $\dot{\Sigma}$ and 
$\dot{g}$, which are enhanced at small $\mu_r$ due to the larger values 
of the coupling constant, are rather comparable at small $x$. The 
present difference between the two cases rather stems from the 
considerably better scale stability of $\dot{g}$ at NLO. 
 
\begin{figure}[h]
\vspace*{1mm}
\centerline{\epsfig{file=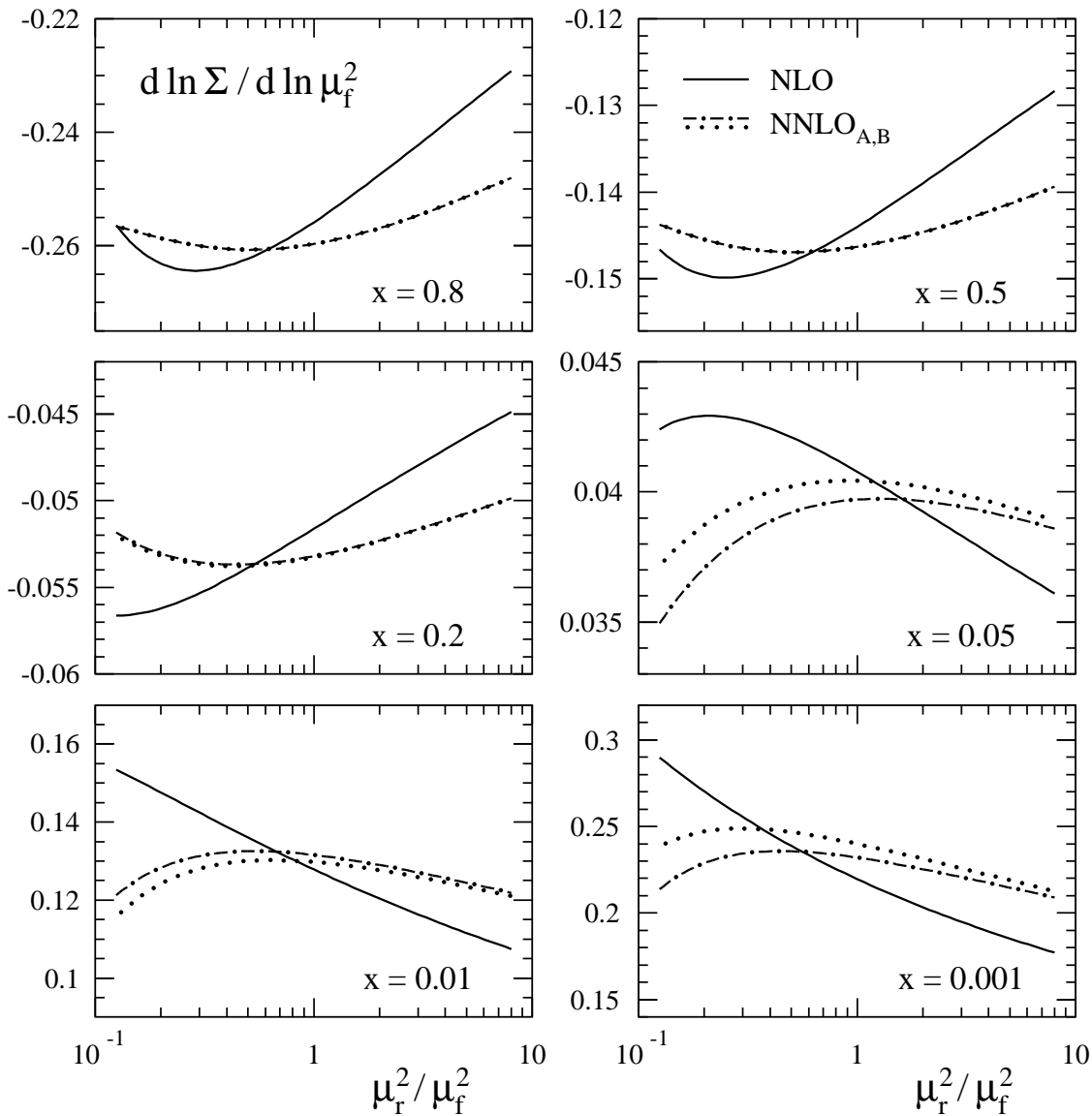,width=15cm,angle=0}}
\vspace*{-1mm}
\caption{The $\mu_r$-dependence of the NLO and NNLO predictions for 
 the singlet quark derivative $\dot{\Sigma} \equiv d \ln \Sigma / 
 d\ln \mu_f^2$ at $\mu_f^2 = \mu_{f,0}^2 \simeq 30 \mbox{ GeV}^2$ 
 for six typical values of $x$.}
\end{figure}

\newpage

\begin{figure}[h]
\vspace*{-1mm}
\centerline{\epsfig{file=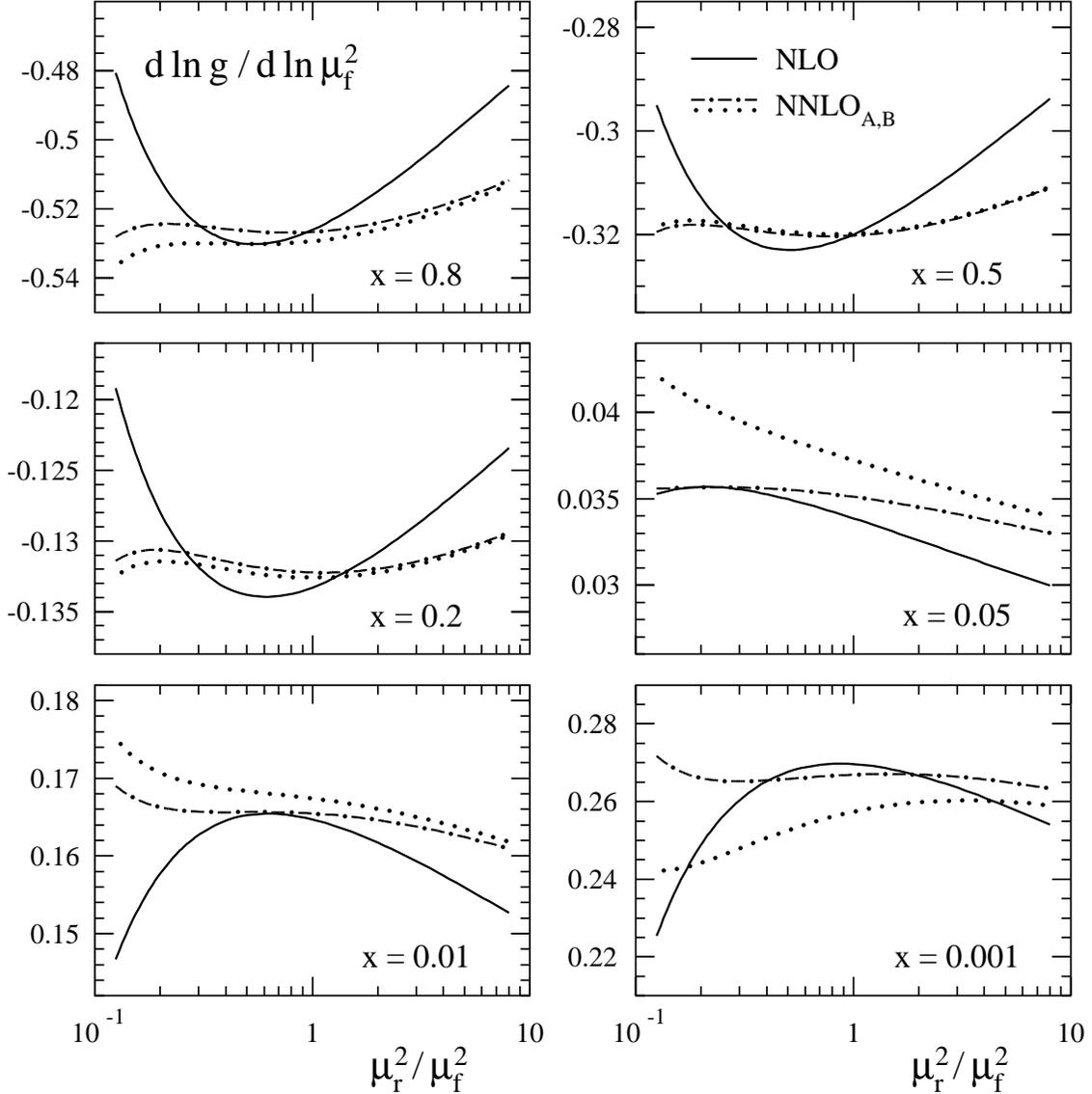,width=15cm,angle=0}}
\vspace{-2mm}
\caption{As Fig.~7, but for the gluon derivative $\dot{g} \equiv d 
 \ln g / d\ln \mu_f^2$. Notice that the scales of the ordinates of the 
 graphs differ within as well as between the two figures.}
\vspace{1mm}
\end{figure}

In Fig.~9 we show the relative renormalization scale uncertainties of 
the singlet quark and gluon evolution, estimated using the smaller 
conventional range $\frac{1}{4}\mu_f^2 \leq \mu_r^2 \leq 4 \mu_f^2$ via
\beq
 \Delta \dot{q} \, \equiv \,
 \frac{\max\, [ \dot{q}(x,\mu_r^2 = \frac{1}{4} \mu_{f}^2 \ldots 4 
 \mu_{f}^2)] - \min\, [\dot{q}(x, \mu_r^2 = \frac{1}{4}\mu_{f}^2 
 \ldots 4 \mu_{f}^2)] }
 { 2\, |\, {\rm average}\, [\dot{q}(x, \mu_r^2 =
 \frac{1}{4}\mu_{f}^2 \ldots 4 \mu_{f}^2)]\, | }
 \:\: , \quad q = \Sigma,\, g \:\: .
\eeq
At large $x$ this estimate yields uncertainties below 2\% and 1\% at 
NNLO for $\dot{\Sigma}$ and $\dot{g}$, respectively, representing 
improvements by more than a factor of three with respect to the
corresponding NLO results. A clear improvement is also found for the 
singlet quark derivative $\dot{\Sigma}$ at small $x$, persisting even 
down to $x=10^{-4}$ at $\mu_f^2 = \mu_{f,0}^2 \simeq 30 \mbox{ GeV}^2$.
For the gluon density, on the other hand, such a small-$x$ 
improvement will only occur at NNLO if $P_{gg}^{(2)}(x)$ is closer 
to the (small) approximation A than to the function B (cf.~Fig.~4).
 
\begin{figure}[thb]
\vspace*{1mm}
\centerline{\epsfig{file=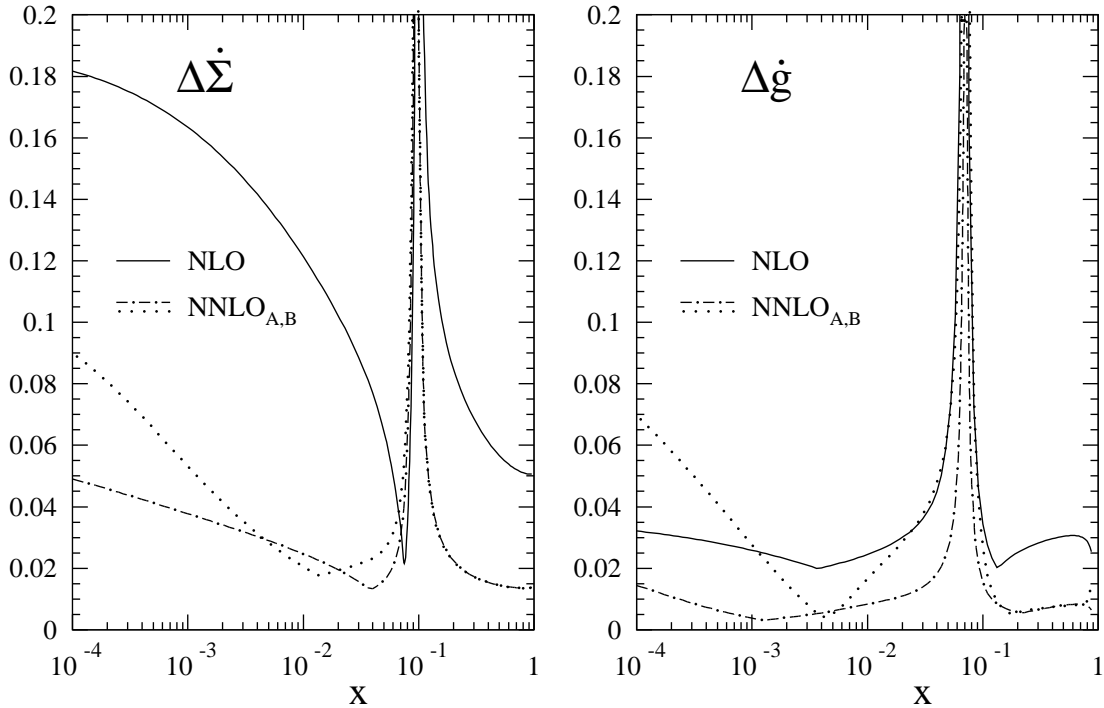,width=15cm,angle=0}}
\vspace{-1mm}
\caption{The relative $\mu_r$-uncertainties of the NLO and NNLO 
 predictions for $\dot{\Sigma}$ and $\dot{g}$ at $\mu_f^2 = \mu_{f,0}^2 
 \simeq 30 \mbox{ GeV}^2$, as estimated by the quantities $\Delta 
 \dot{\Sigma}$ and $\Delta \dot{g}$ defined in Eq.~(5.3). The spikes 
 close to $x = 0.1$ derive from zeros in the denominator of Eq.~(5.3).}
\end{figure}

We now turn to the perturbative expansion (2.16) for the singlet 
structure function $F_{2,S}$ (henceforth simply denoted by $F_2$) and 
its $Q^2$--derivative. 
The corresponding figures below refer to $Q^2 = \mu_{f,0}^2 \simeq 30 
\mbox{ GeV}^2$ and the input (5.1) and (5.2).  For brevity the results 
will be displayed for $\mu_r = \mu_f \equiv \mu$ only.  As above, the 
present NNLO uncertainties due to the incomplete information on the 
three-loop splitting functions are represented by the bands spanned by 
the NNLO$_A$ and NNLO$_B$ approximations.

The LO, NLO and NNLO results are compared in 
Fig.~10 for the standard scale choice $\mu^2 = Q^2$.  Here the NLO and 
NNLO corrections to $F_2$, shown in the left part of the figure, derive 
from the coefficient functions only.  The positive corrections at large 
$x$ stem from the non-singlet parts of the quark coefficient functions 
$c_{2,q}^{(n)}$, $n=1,2\,$, which receive large contributions 
$\,\sim [\ln^k (1-x)/(1-x)]_+$, $k = 0,\,\ldots ,\, 2n\! -\! 1$, from 
soft-gluon emissions.  At $x = 0.8$, for example, the NNLO term adds 
15\% to the NLO result, which in turn exceeds the LO expression 
$\frac{5}{18}\,\Sigma(x,Q^2)$ by 45\%.  
The sizeable negative NNLO corrections at small $x$, on the other 
hand, are mainly due to the gluon contribution.  In this region 
$c_{2,g}^{(2)}(x)$ is dominated by the $1/x$ term 
arising from $t$-channel soft gluon exchange.  
Due to the convolution with the gluon density, however, the other 
contributions to $c_{2,g}^{(2)}$ are as important, as already pointed 
out in refs.~\cite{ZvN2,talk}.  For our input (5.1) the total NNLO 
effect at $10^{-4}\! <\!x\! <\! 10^{-2}$ amounts to about 40\% of 
that without the (positive) $1/x$ terms.

\begin{figure}[thb]
\vspace*{1mm}
\centerline{\epsfig{file=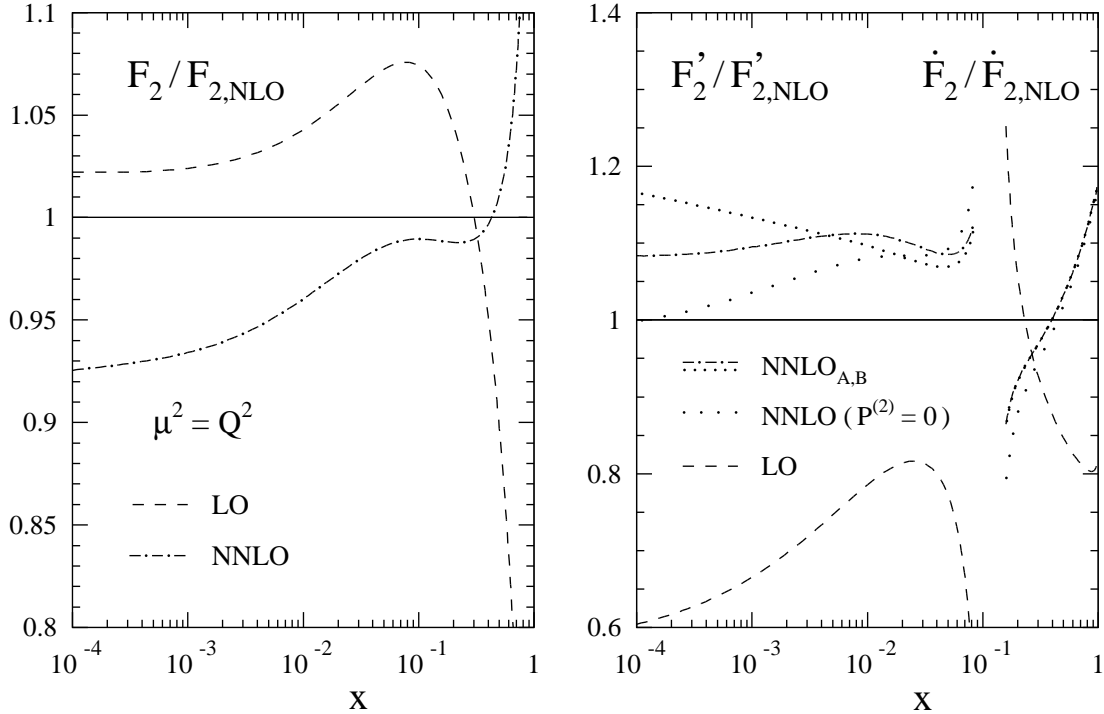,width=15cm,angle=0}}
\vspace{-1mm}
\caption{Comparison of the LO, NLO and NNLO results for the singlet 
 structure function $F_{2}$ (left part), its linear derivative 
 $F_2^{\,\prime}\equiv dF_2/ d\ln Q^2$ (right part, small $x$), and its
 logarithmic derivative $\dot{F}_2 \equiv d \ln F_2/ d\ln Q^2$ (right 
 part, large $x$) at $\mu_r^2 =\mu_f^2 \equiv\mu^2 = Q^2$. All results 
 refer to the parton densities of Eq.~(5.1) and $\alpha_s = 0.2$.}
\end{figure}

Like the evolution of the singlet quark density, the $Q^2$-derivative 
of $F_2$ shown in the right part of Fig.~10 is dominated by the quark 
contribution at large $x$, $x > 0.3$, and by the gluon contribution at 
small $x$, $x < 0.03$.  It is thus convenient to consider the 
logarithmic derivative $\dot{F}_2 \equiv d \ln F_2/ d\ln Q^2$ in the 
former $x$-range, while in the latter region the linear derivative 
$F_2^{\,\prime} \equiv dF_2/ d\ln Q^2$ is more appropriate.  Due to the
effects of the quark coefficient functions discussed above, the LO and 
NNLO corrections to $\dot{F}_2$ at large $x$ are considerably larger 
than their counterparts for $\dot{\Sigma}$ illustrated in Fig.~5.  The
NNLO corrections rise from 3\% at $x = 0.5$ to 11\% at $x = 0.8$, the 
corresponding NLO contributions amount to 18\% and 24\% of the LO 
results.  The small positive gluon contribution to $\dot{F}_2$ 
receives large corrections as well. At $x = 0.5$, for instance, it 
reaches 1.7\%, 3.4\% and 4.8\% of the total $|\dot{F}_2|$ at LO, NLO, 
and NNLO, respectively, in the latter approximation falling below 1\%
only at $x \simeq 0.7$.
$F_2^{\,\prime}$ exhibits a NNLO effect of about 10\% at $10^{-4} \lsim 
x \lsim 0.05$, while the corresponding NLO/LO ratio rises from about 
1.2 at $x = 0.03$ to 1.7 at $x = 10^{-4}$. This rather constant NNLO 
correction combines the effects of the coefficient functions (which 
dominate for $x > 0.01$, but decrease below) 
 and the three-loop splitting functions 
(the impact of which rises towards very small $x$, cf.~Fig.~5). The 
NNLO uncertainties due to the incomplete information on 
$P^{(2)}_{ij}(x)$ are very similar to those for the quark evolution, 
i.e., they reach about $\pm 2\% $ at $x = 10^{-3}$ and then sharply
increase towards $x \ra 0$.

The scale dependence of $F_2$ and of its $Q^2$-derivatives is 
illustrated in Figs.~11 and 12 for the same six $x$-values as in 
Figs.~7 and 8. Analogous to those figures $\mu$ is varied over the 
range $\frac{1}{8}\, Q^2 \leq\mu^2 \leq 8\,Q^2$. The resulting 
estimates for the relative scale uncertainties
\beq
 \Delta f \, \equiv \,
 \frac{\max\, [ f(x,\mu^2 = \frac{1}{4} Q^2 \ldots 4 Q^2)] 
      - \min\, [f(x, \mu^2 = \frac{1}{4}Q^2 \ldots 4 Q^2)] }
 { 2\, |\, {\rm average}\, [f(x, \mu^2 =
 \frac{1}{4}Q^2 \ldots 4 Q^2)]\, | }
 \:\: , \quad f = F_2, \, F_2^{\,\prime},\, \dot{F}_2 \:\: ,
\eeq
are presented in Fig.~13. Except for the intermediate $x$-region $0.03 
\leq x \leq 0.3$ (where $F_2$ is quite stable already at NLO and 
$\dot{F}_2$ is small) the NNLO results represent an improvement by a 
factor two or more at $x \gsim 10^{-4}$ for $F_2$ and $10^{-3} \lsim
x \lsim 0.7$ for $F_2^{\,\prime}$. 

\begin{figure}[htbp]
\vspace*{3mm}
\centerline{\epsfig{file=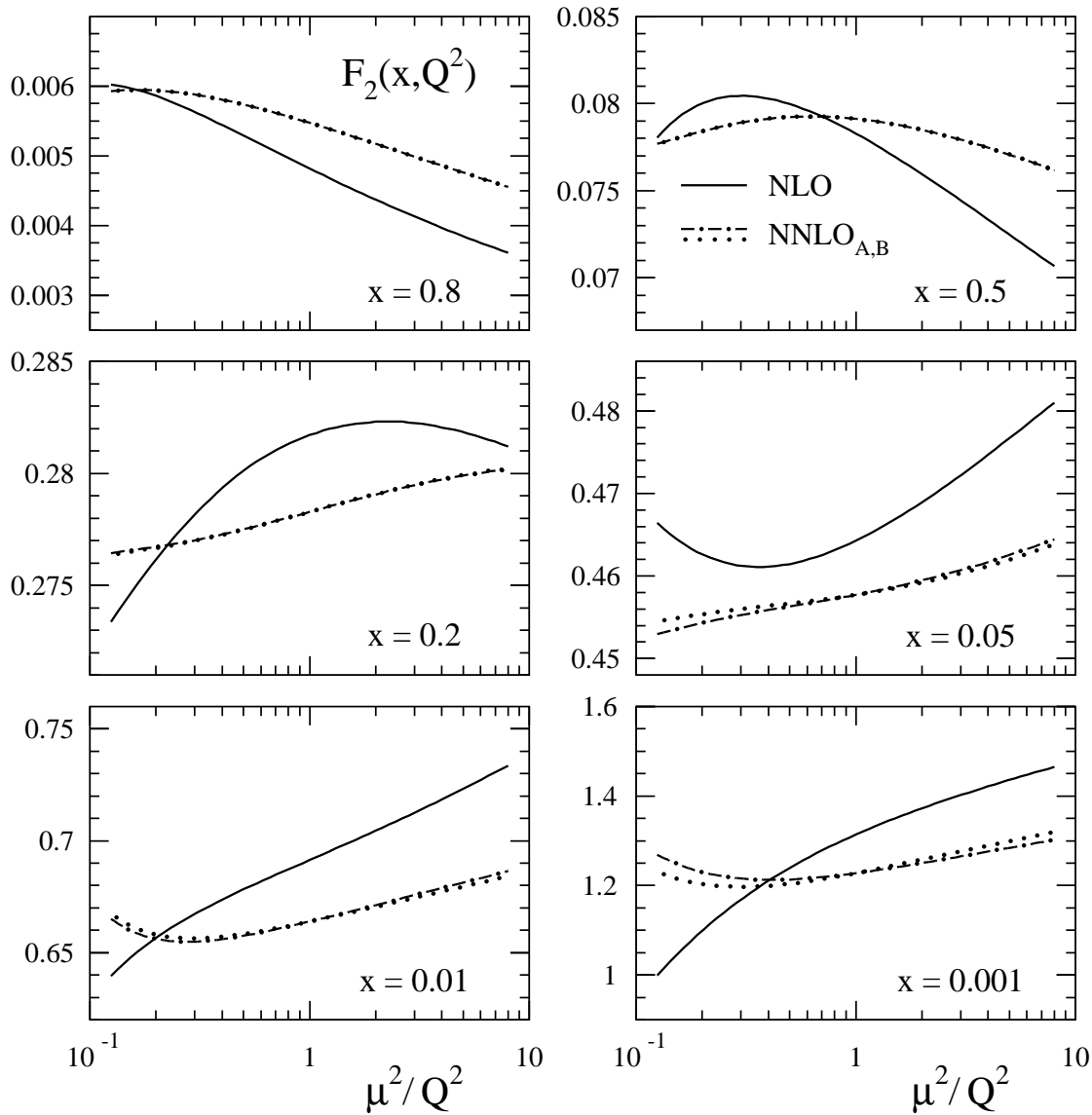,width=15cm,angle=0}}
\vspace*{-1mm}
\caption{The dependence of the NLO and NNLO predictions for the singlet
 structure function $F_2(x,Q^2\! =\! \mu_{f,0}^2)$ on the scale $\mu$ 
 for six representative values of $x$.}
\end{figure}

\newpage

\begin{figure}[htbp]
\centerline{\epsfig{file=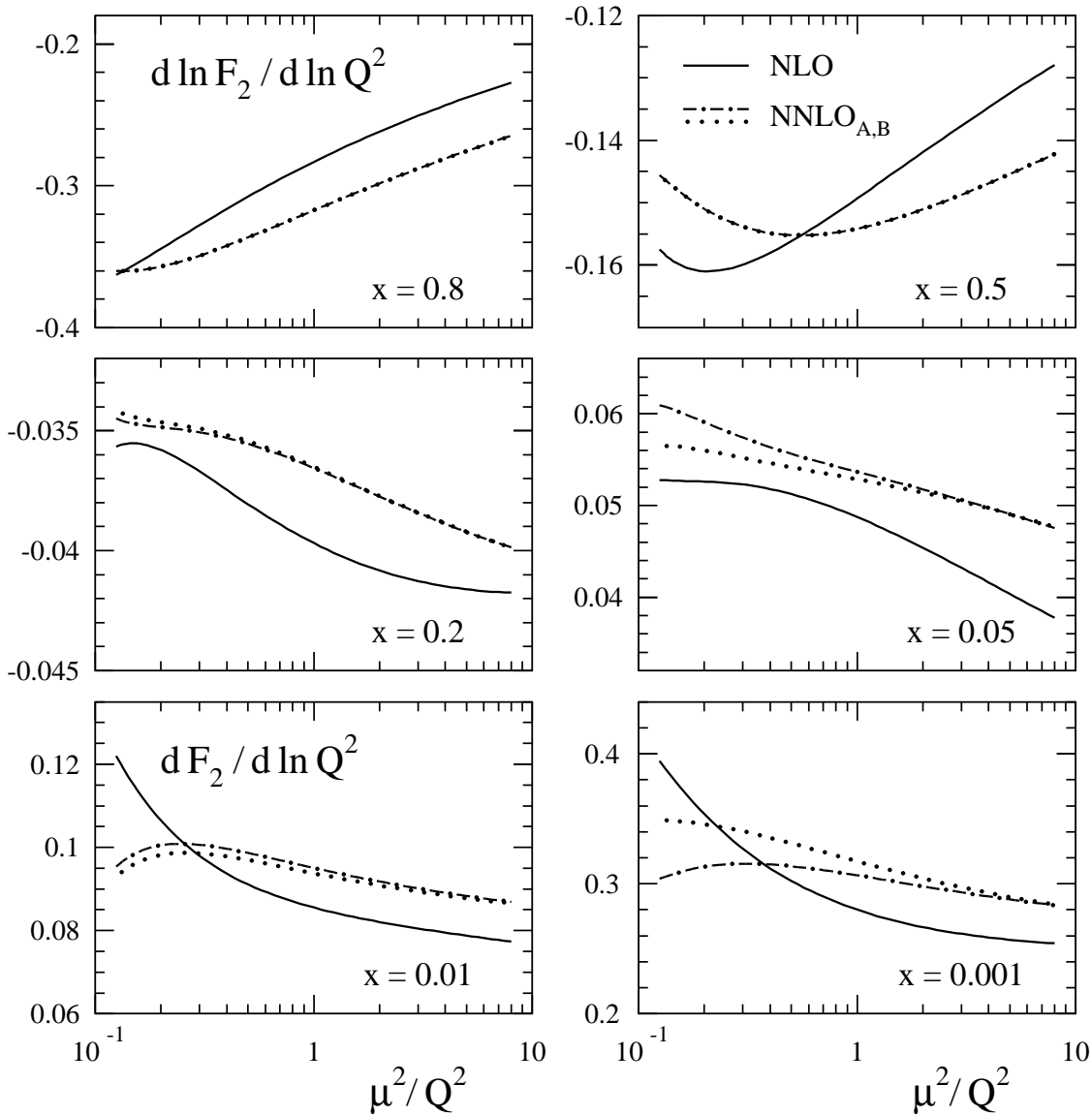,width=15cm,angle=0}}
\vspace{-2.5mm}
\caption{As Fig.~11, but for the logarithmic derivative $\dot{F}_2$
 (upper four $x$-bins) and the linear derivative $F_2^{\,\prime}$ 
 (lower two values of $x$).}
\end{figure}

\begin{figure}[t]
\vspace*{1mm}
\centerline{\epsfig{file=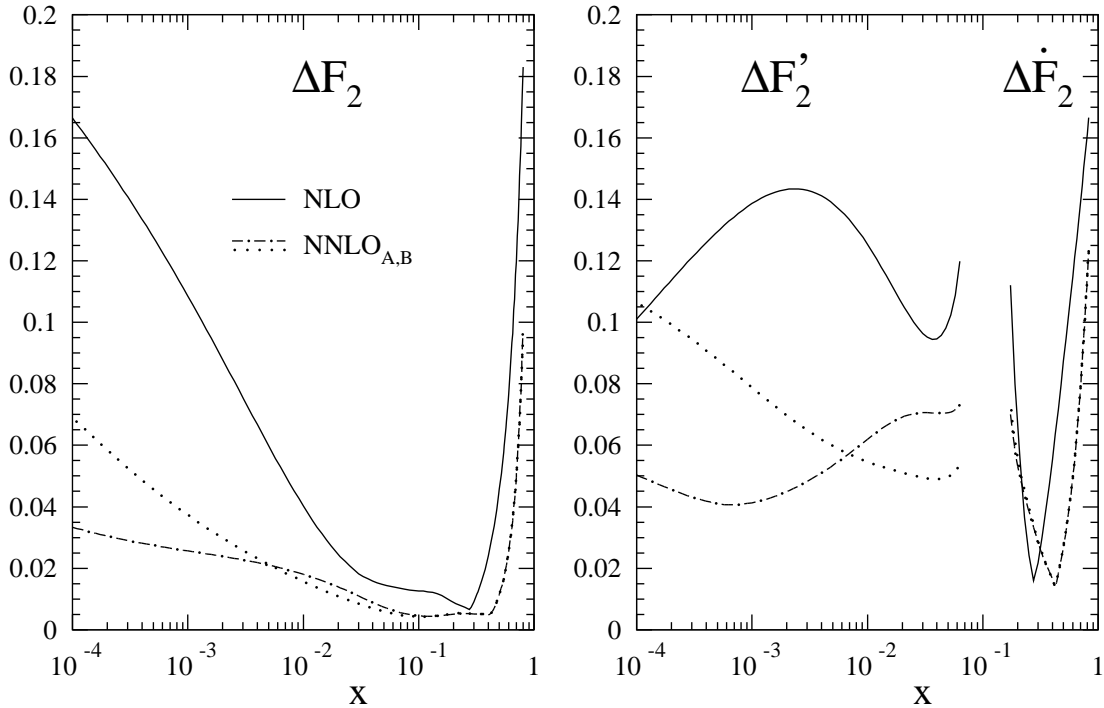,width=15cm,angle=0}}
\vspace{-1mm}
\caption{The relative scale uncertainties of the NLO and NNLO 
 predictions for $F_2(x,Q^2)$ and its linear (small $x$) and logarithmic
 (large $x$) $Q^2$-derivatives at $Q^2 \simeq 30 \mbox{ GeV}^2$, as 
 estimated by the quantities $\Delta F_2$, $\Delta F_2^{\,\prime}$ and 
 $\Delta \dot{F}_2$ defined in Eq.~(5.4).}
\end{figure}

It is also interesting to consider separately the dependence of $F_2$
and its derivatives on the renormalization scale $\mu_r$ (keeping 
$\mu_f^2 = Q^2 = \mu_{f,0}^2$) and on the factorization scale~%
$\mu_f$ (keeping $\mu_r^2 = Q^2 = \mu_{f,0}^2$).  For $F_2$ the 
resulting variations are comparable over the full $x$-range at NNLO.  
At $x = 0.8$, for example, they are both about half as large as the 
$\mu$-depen\-dence shown in Fig.~11.  It is worth noting that the 
$\mu_r$-variation of $F_2$ at small-$x$ worsens by the transition from 
NLO to NNLO, despite the fact that a renormalization scale logarithm 
enters for the first time only at NNLO, see Eqs.~(2.16) and (2.18). 
This is due to the large contribution of $c_{2,g}^{(2)}$ shown in 
Fig.~10. When both scales are varied, however, this deterioration is 
overcompensated by the clear improvement of the $\mu_f$-dependence. 
For the $Q^2$-derivatives of Fig.~12 the effect of $\mu_f$ is much 
smaller than that of $\mu_r$, except for the NLO case at small $x$ 
(where the large scale dependence of $\dot{\Sigma}$ of Fig.~9 is 
relevant).  At NNLO the dependence on $\mu_f$ is relatively largest at 
very large $x$, but even at $x = 0.8$ it does not exceed a quarter of 
the $\mu$-variation shown in Fig.~12.

So far the NNLO effects have been discussed for the fixed initial 
parton densities $\Sigma (x,\mu_{f,0}^2)$ and $g(x,\mu_{f,0}^2)$ of 
Eq.~(5.1) and the coupling constant $\alpha_s(\mu_{f,0}^2) = 0.2$. We 
now finally address the impact of the NNLO terms on the determination 
of these perturbatively incalculable inputs from data on $F_2$. For a 
simple estimate of the resulting difference between the NLO and the 
NNLO singlet parton densities and their respective theoretical 
uncertainties, we employ as `data' the values of $F_2(x,Q^2)$ and 
$F_2^{\,\prime}(x,Q^2)$ for $Q^2 = \mu_{f,0}^2 \simeq 30 \mbox{ GeV}^2$ 
at the six $x$-values of Figs.~11 and 12, supplemented by $x= 10^{-4}$. 
I.e., the parameters of the input (5.1) (including a $(1+Ax)$ 
interpolation term for the gluon density, but imposing the momentum sum 
rule) are fitted at NNLO for $\mu^2 = Q^2$ to the corresponding NLO 
results. The errors bands due to the scale dependence are then 
determined by repeating the NLO and NNLO fits to their respective 
$\mu^2\! =\! Q^2$ central values of $F_2$ and $F_2^{\,\prime}$ for 
$\mu^2 / Q^2 =$ 0.25, 0.5, 2 and 4.
 
A critical point in analyses of $F_{2,S}$ is the correlation between 
$\alpha_s$ and the gluon density, which tends to strongly increase the 
$\mu$-variation of the fitted values of $\alpha_s$.  Indeed, under the 
present conditions this variation is enhanced, in both NLO and NNLO, by 
about a factor of two with respect to an analogous analysis of the 
non-singlet structure function $F_{2,\rm NS}$, thus preventing an 
accurate determination of $\alpha_s$ from $F_{2,S}$ alone.
In order to keep the scale variation of $\alpha_s$ at a level realistic 
for analyses of data on electromagnetic DIS (where both 
$F_{2,\rm NS}^{\, p}$ and $F_{2,\rm S}^{\, p}$ have been accurately 
measured), we therefore take over the values for $\alpha_s
(\mu_{f,0}^2)$ from our previous non-singlet analysis \cite{NV1}: 
$\alpha_s(\mu_{f,0}^2)$ = 0.188, 0.200 and 0.220, respectively, for 
$\mu^2 = 0.25\, Q^2$, $Q^2$ and $4\, Q^2$ and NLO. The corresponding 
NNLO values read 0.190, 0.194 and 0.202.  It is clear from our 
discussion above that these scale uncertainties of $\alpha_s$ are 
almost entirely due to the variation of the renormalization scale.

The resulting NNLO central values and NLO and NNLO scale-uncertainty
bands for $\Sigma (x,\mu_{f,0}^2)$ and $g(x,\mu_{f,0}^2)$ are presented
in Fig.~14. The shift of $\Sigma_{\rm NNLO}$ with respect to 
$\Sigma_{\rm NLO}$ -- an enhancement between 3\% and 7\% at small $x$,
and a sizeable decrease at very large $x$ reaching 13\% at $x = 0.8$ -- 
rather directly reflects the results for $F_2$ shown in Fig.~10.
Likewise the reduction of the scale dependence to $\pm 2\% $ or less
for $10^{-3}\!< x < 0.5$ closely follows the pattern of $\Delta F_2$ in 
Fig.~13. The corresponding results for the gluon density, on the other 
hand, are considerably affected by the values of $\alpha_s$. In fact,
the theoretical error band for $g_{\rm NNLO}^{\,}$ does not exceed 2\% 
for $3\!\cdot\! 10^{-3}\! < x < 0.2$. For $x < 10^{-3}$ the present 
uncertainties of the three-loop splitting functions prevent an 
improvement on the rather accurate NLO determination of the gluon 
density.

\begin{figure}[t]
\vspace*{1mm}
\centerline{\epsfig{file=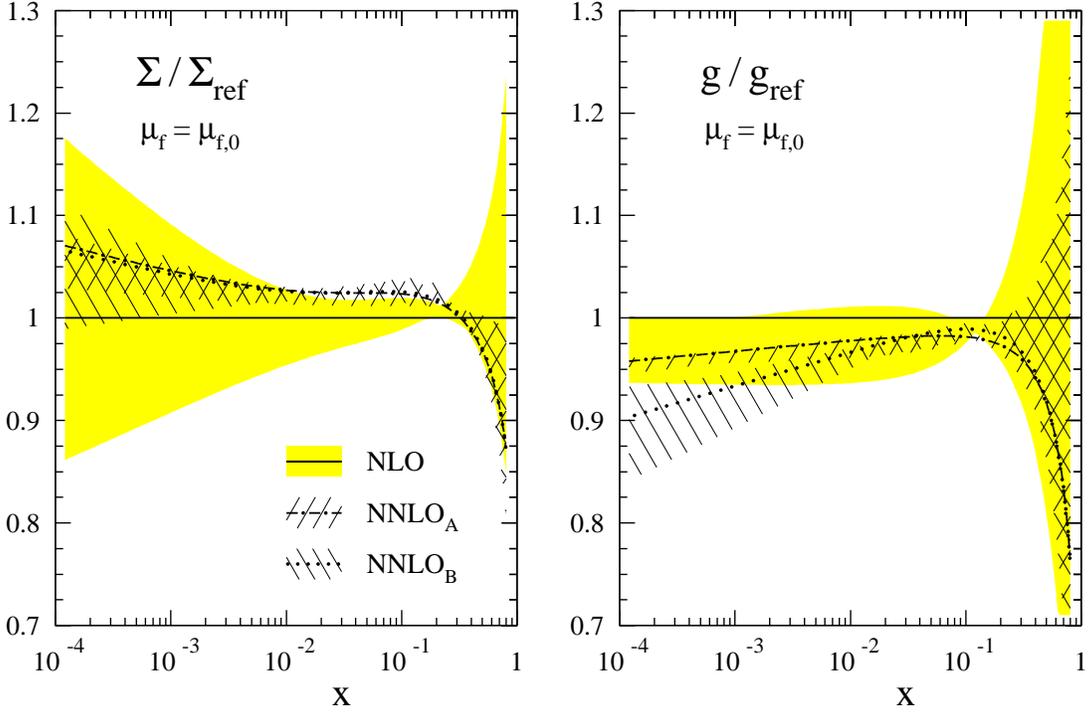,width=15cm,angle=0}}
\vspace{-1mm}
\caption{Comparison of the NLO and NNLO singlet quark and gluon 
 densities and their scale-uncertainty bands as obtained by fitting 
 $F_2$ and $F_2^{\,\prime}$ at $Q^2 = \mu_{f,0}^2 \simeq 30 
 \mbox{ GeV}^2$ for $10^{-4} \leq x \leq 0.8$. The range of scales used 
 for the error bands is $0.25\, Q^2 \leq \mu^2 \leq 4\, Q^2$.}
\end{figure}
%
\section{Summary}
%
%
We have investigated the effect of the NNLO perturbative QCD 
corrections on the evolution of the unpolarized flavour-singlet 
quark and gluon densities, $\Sigma(x,\mu_f^2)$ and $g(x,\mu_f^2)$,
and on the most important singlet structure function, $F_{2,S}
(x,Q^2)$.  Our main new ingredients are approximate expressions for 
the $x$-dependence $P_{ij}^{(2)}(x)$ of the singlet three-loop 
splitting functions, including quantitative estimates of their residual 
uncertainties.  These parametrizations have been derived from the 
lowest four even-integer moments $P_{ij}^{(2)}(N)$, $N = 2$, 4, 6 and 
8, calculated in ref.\ \cite{spfm2}, supplemented by the results of 
refs.\ \cite{CH94,FL98} for the leading small-$x$ terms $\propto (1/x) 
\ln x$ of $P_{qq}^{(2)}$, $P_{qg}^{(2)}$ and $P_{gg}^{(2)}$. 
Consequently the differences between our parametrizations illustrated 
in Figs.~3 and 4 oscillate at large $x$, as the difference between any 
two approximations obviously has four vanishing moments.
At $x\,\lsim\, 10^{-2}$ the differences increase, because the  
$(1/x) \ln x$ terms unfortunately do not sufficiently dominate over 
the unknown less singular contributions like $1/x$ in the 
experimentally accessible region of small $x$, see also refs.\
\cite{BNRV,BvN}.

Clearly our parametrizations represent only a temporary `solution' of 
the problem of the three-loop splitting functions, which will be 
superseded sooner or later by an exact calculation. However, for two 
reasons the present approach is, over a wide region of $x$, much more 
effective than a brief look at Figs.~3 and 4 might suggest. First of 
all the splitting functions enter physical quantities only via 
convolutions with nonperturbative initial distributions, which smoothen 
out the above-mentioned oscillations to a large extent. While this 
mechanisms leads to especially small uncertainties at $x\,\gsim\, 0.1$, 
the convolutions extend the relevance of the large-$x$ constraints 
\cite{spfm2} deep into the small-$x$ region as exemplified in the 
discussion of Fig.~2. 
The second reason is that the series expansion for the evolution of the 
parton densities is very well converging --- except, possibly, at 
$x<10^{-3}$. Choosing for definiteness $\alpha_s = 0.2$ for the strong 
coupling constant (this corresponds to a scale between about 20 GeV$^2$
and 50 GeV$^2$, depending on the precise value of $\alpha_s(M_Z^{2})
^{\,}$) the NNLO effect on the evolution of $\Sigma(x,\mu_f^2)$ and 
$g(x,\mu_f^2)$ amounts to less than 2\% and 1\%, respectively, at 
$x\,\gsim\, 0.2$.  Hence the net uncertainties due to the incomplete 
information on the $P_{ij}^{(2)}(x)$ are absolutely negligible in this 
region as shown in Figs.~5 and 6.  Of course, these uncertainties rise 
towards small $x$, but they exceed $\pm 2\%$ only below 
$x \simeq 10^{-3}$ (or a few times this number, if much lower scales 
are involved), leaving a comfortably large region of safe applicability 
of our results to processes at the {\sc Tevatron} and the LHC.

At $x\,\gsim\, 0.2$ the NNLO renormalization-scale variation (for the 
conventional interval $\frac{1}{4}\, \mu_f^2 \leq \mu_r^2 \leq 4\, 
\mu_f^2$) amounts to less than 2\% and 1\%, respectively, for $d \Sigma 
/ d\ln \mu_f^2$ and $dg / d\ln \mu_f^2$ at our reference point 
$\mu_{f}^2\simeq 30$ GeV$^2$. The corresponding numbers at $x= 10^{-3}$ 
read 5\% and 3\%.  Except for the gluon evolution at this latter value
of $x$, these results represent an improvement on the NLO evolution 
by a factor of three or more.  Taking into account also the rapid 
convergence at $\mu_r = \mu_f$ and our corresponding findings for the 
non-singlet sector in ref.\ \cite{NV1}, we expect that terms beyond 
NNLO will affect the parton evolution by less than 1\% at large $x$ and 
less than 2\% down to $x \simeq 10^{-3}$ for $\alpha_s \simeq 0.2$. 

Due to the additional effect of the two-loop coefficient functions
\cite{ZvN1,ZvN2}, the structure function $F_{2,S}$ and its scaling 
violations receive considerably larger NNLO corrections at $x>10^{-2}$.
In fact, keeping only the coefficient functions (and omitting the 
three-loop splitting functions $P_{ij}^{(2)}(x)$ altogether) forms 
quite a good approximation in this $x$-range. 
As shown in Fig.~10, the NNLO corrections for both $F_2$ and its 
$Q^2$-derivative are particularly large at very large $x$, e.g., 15\%
for $F_{2,S}$ and 11\% for $d\ln F_{2,S} / d\ln Q^2$ at $x = 0.8$ for
our reference scale $Q^2 \simeq 30\mbox{ GeV}^2$.  This is an effect of 
the large soft-gluon terms in the quark coefficient functions which are 
not special to the singlet case considered here.  While the gluon 
contributions dominates the sizeable (up to about 7\%) NNLO corrections 
at small $x$, their effect is suppressed at large $x$, especially for 
the absolute size of $F_{2,S}$.  It is worth noting, however, that that 
the gluon contribution to $dF_{2,S} / d\ln Q^2$ at $x = 0.5$ still 
amounts to 5\% at NNLO (40\% more than at NLO), an effect large enough 
to jeopardize analyses which apply a purely non-singlet formalism to 
the data on the proton structure function $F_2^{\, p}(x,Q^2)$ in the 
region $x > 0.3$. 
The accuracies of both $F_2$ and its scaling violations, as estimated 
by the scale dependence, are considerably improved (by more than a 
factor of two over a wide region) by the inclusion of the NNLO terms.  
Consequently the same applies to the theoretical accuracy of 
determinations of the singlet quark and gluon densities from data on 
$F_{2,S}$ and $dF_{2,S} / d\ln Q^2$ at $Q^2 \simeq 30\mbox{ GeV}^2$ 
illustrated in Fig.~14: uncertainties of less than 2\% from the 
truncation of the perturbation series are obtained for the quark 
density at $10^{-3}\! <\! x\! <\!  0.5$ and for for the gluon density 
at $3\cdot 10^{-3}\! <\! x \! <\! 0.2$.

Our results of ref.~\cite{NV1} and the present paper complete, if only 
approximately at $x\! >\! 10^{-3}$, the theoretical prerequisites for 
NNLO analyses of structure functions in DIS and total cross sections 
of Drell-Yan processes.  
Further progress at large $x$ ---  especially a further improvement on 
the theoretical accuracy of NNLO $\alpha_s$-determinations from 
structure functions \cite{KPS,SY99,NV1} --- can be obtained by 
including the $O(\alpha_s^3)$ coefficient functions, particularly in 
the non-singlet sector dominating the extraction of $\alpha_s$. In 
fact, results on these functions are available from the fixed-moment 
calculations of refs.~\cite{spfm1,spfm2} and from soft-gluon 
resummation \cite{sglue,av99}.  We will address this issue in a 
forthcoming publication. 
Progress towards the important HERA small-$x$ region of $x \lsim 
10^{-3}$ at moderate to low $Q^2$, however, definitely requires the 
full calculation of the three-loop splitting functions.

{\sc Fortran} subroutines of our parametrizations of 
$c_{a,i}^{(2)}(x)$ ($a = 2, L$; $i = q,g$) and $P_{ij}^{(2)}(x)$ can 
be obtained via email to neerven@lorentz.leidenuniv.nl or
avogt@lorentz.leidenuniv.nl.
%
%
\newpage
\section*{Acknowledgment}
%
%
This work has been supported by the European Community TMR research
network `Quantum Chromodynamics and the Deep Structure of Elementary 
Particles' under contract No.~FMRX--CT98--0194.

\vspace{1cm}
 
%
%
\section*{Appendix: The singlet coefficient functions in 
          {\boldmath $N$}-space}
\setcounter{equation}{0}
\renewcommand{\theequation}{A.\arabic{equation}}
%
%
The Mellin transforms (2.12) of the parametrizations (3.3)--(3.6) of 
the two-loop singlet coefficient functions for $F_L$ and $F_2$ are 
given in terms of the integer-$N$ sums $S_l(N)$ and their complex-$N$
analytic continuations
\begin{equation}
  S_{l} \equiv S_{l}(N) = \sum_{k=1}^N \frac{1}{k^l}
  = \zeta (l) - \frac{(-1)^l}{(l-1)!} \,\psi^{(l-1)}(N\! +\! 1)
  \:\: .
\end{equation}
Here $\zeta (1) $ represents the Euler--Mascheroni constant, and 
$\zeta (l\! >\! 1)$ Riemann's $\zeta$-func\-tion. The $l\,$th 
logarithmic derivative $\psi^{(l-1)}$ of the $\Gamma$-function can be 
readily evaluated using the asymptotic expansion for ${\rm Re}\, N>10$ 
together with the functional equation.
 
\noindent
The moments (2.12) of the pure singlet coefficient functions (3.3) and 
(3.4) are given by
\begin{eqnarray}
   c_{L,{\rm PS}}^{(2)}(N) \!
   &=& N_f \, \Bigg\{ 
       \bigg( - \frac{15.94}{N} + \frac{37.092}{N+1} 
              - \frac{26.364}{N+2} + \frac{5.212}{N+3} \bigg) S_1
       \nonumber \\[0.5mm]
   & & \mbox{}  
       - \frac{2.370}{N-1}
       + \frac{0.842}{N^3} - \frac{28.09}{N^2} + \frac{26.38}{N}
       + \frac{3.040}{(N+1)^3} + \frac{65.182}{(N+1)^2} 
       \nonumber \\[0.5mm]
   & & \mbox{}  
       - \frac{88.678}{N+1}
       - \frac{26.364}{(N+2)^2} + \frac{91.756}{N+2}
       + \frac{5.212}{(N+3)^2} - \frac{27.088}{N+3}
        \Bigg\}
\end{eqnarray}
and 
\begin{eqnarray}
   c_{2,{\rm PS}}^{(2)}(N) \!
   &=& N_f \, \Bigg\{ 
       \bigg(\frac{49.702}{N} - \frac{27.802}{N+1} \bigg) S_3
       + \bigg(\frac{49.5}{N^2} + \frac{30.23}{N} 
         - \frac{27.903}{(N+1)^2} \bigg) S_2 
       \nonumber \\[0.5mm]
   & & \mbox{}+ 0.101 \bigg( \frac{1}{N} - \frac{1}{N+1} \bigg) 
      \Big(3\, S_2 S_1 + S_1^3 \Big)
       + \bigg( \frac{49.5}{N^3} + \frac{30.23}{N^2} 
      - \frac{28.206}{(N+1)^{3}} \bigg) S_1
       \nonumber \\[0.5mm]
   & & \mbox{}+ \frac{5.290}{N-1} - \frac{25.86}{N^4} 
       - \frac{4.172}{N^3} - \frac{121.205}{N^2} - \frac{114.519}{N} 
       -  \frac{83.406}{(N+1)^4}
       \nonumber \\[0.5mm]
  & & \mbox{}+ \frac{45.4003}{(N+1)^2} + \frac{33.1769}{N+1} \Bigg\}
      \:\: .
\end{eqnarray}
The corresponding expressions for the gluon coefficient functions (3.5)
and (3.6) read
\begin{eqnarray}
   c_{L,g}^{(2)}(N) \!
   &=& N_f \, \Bigg\{
       \bigg( \frac{94.74}{N} + \frac{1017.06}{N+1} + \frac{49.20}{N+2} 
              \bigg) S_2
     + \bigg( \frac{94.74}{N} - \frac{143.94}{N+1} + \frac{49.20}{N+2} 
              \bigg) S_1^2
       \nonumber \\[0.5mm]
   & & \mbox{}
     + \bigg( -\frac{864.8}{N} + \frac{873.12}{(N+1)^2} 
              + \frac{963.2}{N+1} 
              + \frac{98.40}{(N+2)^2} - \frac{98.40}{N+2} \bigg) S_1
       \nonumber \\[0.5mm]
   & & \mbox{}
       - \frac{5.333}{N-1}
       - \frac{39.66}{N^2} + \frac{5.333}{N}
       + \frac{2154.24}{(N+1)^3} + \frac{1002.86}{(N+1)^2}
       - \frac{1909.768}{N+1} 
       \nonumber \\[0.5mm]
   & & \mbox{}
       - \frac{98.40}{(N+2)^3} + \frac{98.40}{(N+2)^2}
        \Bigg\}
\end{eqnarray}
and 
\begin{eqnarray}
   c_{2,g}^{(2)}(N) \!
   &=& N_f \, \Bigg\{ 
       \bigg( \frac{2760.11}{N} + \frac{418.8}{N+1} \bigg) S_3
       + \bigg( \frac{2320}{ N^2} - \frac{24}{N} + \frac{628.2}{(N+1)^2}
       \bigg) S_2 
       \nonumber \\[0.5mm]
   & & \mbox{}+ \bigg( \frac{1096.07}{N} + \frac{628.2}{N+1} \bigg) 
       S_2 S_1 
       + \bigg( \frac{871.8}{N^2} - \frac{24}{N} + \frac{628.2}{(N+1)^2}
        \bigg) S_1^2
       \\
   & & \mbox{}- \bigg( \frac{1494}{N\! -\! 1} - \frac{1448.20}{N^3} 
       +  \frac{1385.12}{N} -  \frac{1256.4}{(N\! +\! 1)^3} \bigg) S_1 
       - \bigg( \frac{215.845}{N} - \frac{209.4}{N\! +\! 1} \bigg) S_1^3
       \nonumber \\[0.5mm]
   & & \mbox{}+ \frac{1505.9}{N-1} -  \frac{31.914}{ N^4} 
       - \frac{118.96}{N^3} - \frac{2097.4}{N^2} - \frac{4938.34}{N} 
       + \frac{1256.4}{ (N+1)^4} - 0.271 \Bigg\} \:\: .
       \nonumber
\end{eqnarray}
 
Let us finally briefly elaborate on the derivation of Eq.~(4.7) from 
Eq.~(4.5).  The exact expression \cite{ZvN1,ZvN2} for the residue of 
the $N=1$ pole of $c_{2,g}^{(2)}(N)$ reads $N_f\,[344/9-16\,\zeta(2)]$.
Recalling that the Mellin transform of $(1/x) \ln x$ is given by
$-1/(N\! -\! 1)^2$, this information combined with the corresponding 
term $16/[3(N\! -\! 1)]$ of the LO splitting function $P_{gq}^{(0)}$ is 
sufficient to obtain Eq.~(4.7) from Eqs.~(4.5) and (4.6). The latter 
relation simply arises since the other contributions to the scheme 
transformation (2.24) do not lead to any $1/(N\! -\! 1)^2$ terms. 

\vspace{1cm}
 
%

%
\end{document}